\documentclass[12pt]{article}
\usepackage{epsfig,amssymb,amsmath,psfrag,subfigure,rotate,color}

\allowdisplaybreaks
\newcommand{\insertfig}[2]{\includegraphics[width=#1cm]{#2}}

\def \be  {\begin{equation}}
\def \ee  {\end{equation}}
\def \ba  {\begin{eqnarray}}
\def \ea  {\end{eqnarray}}
\def \baa {\begin{eqnarray*}}
\def \eaa {\end{eqnarray*}}
\def \lab #1 {\label{#1}}

\newcommand\re[1]{(\ref{#1})}
\def\d{\hbox{{d}\kern-.20em\hbox{l}}}

\def \matrix #1 {\left(\begin{array}{cc} #1 \end{array}\right)}

\newcommand \ket [1] {|{#1}\rangle}
\newcommand \bra [1] {\langle {#1}|}
\newcommand{\bit}[1]{\mbox{\boldmath$#1$}}

\newcommand{\ft}[2]{{\textstyle\frac{#1}{#2}}}
\begin{document}

\begin{titlepage}

\thispagestyle{empty}

\vspace*{3cm}

\centerline{\large \bf On factorization of multiparticle pentagons}
\vspace*{1cm}

\centerline{\sc A.V.~Belitsky}

\vspace{10mm}

\centerline{\it Department of Physics, Arizona State University}
\centerline{\it Tempe, AZ 85287-1504, USA}

\vspace{2cm}

\centerline{\bf Abstract}

\vspace{5mm}

We address the near-collinear expansion of multiparticle NMHV amplitudes, namely, the heptagon and octagons in the dual language of null
polygonal super Wilson loops. In particular, we verify multiparticle factorization of charged pentagon transitions in terms of pentagons for single 
flux-tube excitations within the framework of refined operator product expansion. We find a perfect agreement with available tree and one-loop data. 

\end{titlepage}

\setcounter{footnote} 0

\newpage

\pagestyle{plain}
\setcounter{page} 1

{
\tableofcontents}

\newpage

\section{Introduction}

The theory of the color flux-tube in planar maximally supersymmetric gauge theory is deeply rooted in the integrability of the model \cite{Beisert:2010jr}. Recently 
an operator product expansion (OPE) in terms of its fundamental excitations was successfully formulated \cite{Alday:2010ku,Basso:2013vsa} to compute the 
expectation value of null polygonal supersymmetric Wilson loop $\mathcal{W}$ to any order of 't Hooft coupling. The superloop is dual to the on-shell scattering 
superamplitude \cite{Alday:2007hr,Drummond:2007cf,Brandhuber:2007yx,CaronHuot:2010ek,Mason:2010yk,Belitsky:2011zm} and thus promises one to provide 
nonperturbatively the complete S-matrix of the super Yang-Mills theory in question. Since the latter is a superconformal theory and thus does not possess 
asymptotic particle states in four-dimensions, one has to deal with regularized and properly subtracted combinations of amplitudes, known as ratio functions 
\cite{Drummond:2008vq,Mason:2009qx}. The refined version of the operator  product expansion approach employs the so-called pentagon transitions 
$P (\psi|\psi^\prime)$ between eigenstates $\psi$ and $\psi^\prime$ of the color flux tube \cite{Basso:2013vsa}. An eigenstate $\psi$ is parametrized by the 
eigenvalues of three generators of the conformal group which yield the energy $E_\psi$, momentum $p_\psi$ and helicity $m_\psi$ of the state. The dispersion 
relation $E = E (p)$ for the latter is conveniently parametrized by the excitation's rapidity $u$, such that $E = E (u)$ and $p = p (u)$. The latter are known to all 
orders in 't Hooft coupling for any excitations propagating on the flux tube \cite{Basso:2010in}.

\begin{figure}[t]
\begin{center}
\mbox{
\begin{picture}(0,220)(60,0)
\put(0,-150){\insertfig{17}{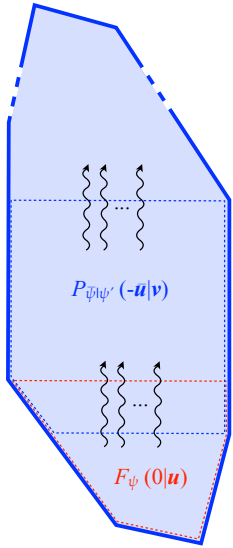}}
\end{picture}
}
\end{center}
\caption{ \label{GenericPolygonInPentagons} A generic polygon tessellated into a sequence of pentagon transitions (shown by dashed contours) 
between eigenstates of the color flux tube. The bottom pentagon is encoded by the form factor for creation of corresponding excitations. Crossing 
relates it to pentagons.}
\end{figure}

An $N$-sided super Wilson loop $\mathcal{W}_N$ in a chosen tessellation is then decomposed in terms of the pentagons as shown in Fig.\ \ref{GenericPolygonInPentagons} 
and reads  \cite{Basso:2013vsa}
\begin{align}
\label{WnOPE}
\mathcal{W}_N = \int d \mu_\psi (\bit{u}) F_\psi (0 | \bit{u}) 
P_{ \bar{\psi} | \psi^\prime} ( -\bar{\bit{u}}| \bit{v}) 
d \mu_{\psi^\prime} (\bit{v})
P_{ \bar{\psi}^\prime | \psi^{\prime\prime}} ( -\bar{\bit{v}}| \bit{w}) 
\dots
\, .
\end{align}
For this polygon, there are $N-6$  intermediate pentagons $P$, which together with the first and last ones, --- dubbed creation/absorption form factors 
$F$ for incoming/outgoing states, --- overlap on $N-5$ intermediate squares. The latter are encoded in the measures $d \mu$ that cumulatively depend on 
$3 (N-5)$ independent conformal cross ratios $\tau_j$, $\sigma_j$ and $\phi_j$. For a given intermediate transition, $d \mu$ gets contribution from an $n$-particle 
state which admits a factorized form
\begin{align}
d \mu_\psi (\bit{u}) \equiv
\prod_{j = 1}^n d \mu_{{\rm p}_j} (u_j)
\, , \qquad
d \mu_{{\rm p}_j} (u_j)
= \frac{d u_j}{2 \pi} \, \mu_{{\rm p}_j} (u_j) \, {\rm e}^{- \tau E_{{\rm p}_j} (u_j) + i \sigma p_{{\rm p}_j} (u_j) + i \phi m_{{\rm p}_j}}
\, .
\end{align}
Here and below, we will associate the first set of the cross ratios $\tau_1, \sigma_1, \phi_1$ with excitation rapidities $\bit{u}$,  the second set $\tau_2, \sigma_2, \phi_2$ 
with  $\bit{v}$, $\tau_3, \sigma_3, \phi_3$ with  $\bit{w}$ etc. Above, the first and last pentagon transitions differ from all intermediate ones by the fact that their initial and 
final states, respectively, correspond to the flux-tube vacuum, however they are related to the rest by mirror transformations \cite{Basso:2013vsa,Basso:2011rc}. 
The chosen conventions were adopted from the integrability-based pentagon framework which one uses to compute all ingredients to any order of perturbation 
theory \cite{Basso:2013vsa,Basso:2013aha,Basso:2014koa,Belitsky:2014sla,Basso:2014nra,Belitsky:2014lta,Basso:2014hfa} from a set of axioms 
\cite{Basso:2013vsa}. Hence, we employed the following notations in Eq.\ \re{WnOPE}: while $\psi$ corresponds to a collection of excitations in particular 
order $\psi = \{ {\rm p}_1, \dots, {\rm p}_n \}$, $\bar\psi$ stands for its reverse $\bar\psi = \{ {\rm p}_n, \dots, {\rm p}_1\}$. The same nomenclature  applies to their 
rapidities associated with the corresponding flux-tube excitations, $\bit{u} = \{ u_1, \dots, u_n \}$ and $\bar{\bit{u}} = \{ u_n, \dots, u_1 \}$, respectively. 

The integrals over the rapidities of the flux-tube excitations in the formula \re{WnOPE} go over specific contours. The latter are the same for both holes and 
gauge fields which run along the real axis (or slightly above it in the complex plane). The most complicated contour $C$ is the one for fermionic excitations 
\cite{Basso:2014koa}. It travels over a two-sheeted Riemann surface with a cut along the interval on the real axis $[-2g, 2g]$: in a nutshell, it is 
conveniently decomposed within perturbation theory into two contours $C = C_{\rm F} \cup C_{\rm f}$, with $C_{\rm F}$ that lies on the so-called large 
fermion sheet and runs just above the real axis, $C_{\rm F} = (-\infty + i 0, + \infty + i 0 )$, the second one is a half-circle closed contour $C_{\rm f}$ on the 
small fermion sheet in the lower half plane of the complex rapidity plane. Having spelled out these explicitly, in what follows, we will not display the integration 
domains explicitly.

The OPE for the hexagon was addressed in great detail in previous studies \cite{Basso:2013vsa,Basso:2013aha,Basso:2014koa,Belitsky:2014sla,Basso:2014nra,Belitsky:2014lta}
and successfully compared with available higher-loop data on scattering amplitudes\footnote{Quite recently similar results were obtained for the symbol of MHV heptagon in Refs.\ 
\cite{Golden:2014xqa,Golden:2014xqf,Golden:2014pua,Drummond:2014ffa} up to three loop order.} 
\cite{DelDuca:2010zg,Goncharov:2010jf,Dixon:2013eka,Dixon:2014voa,Dixon:2011nj,Dixon:2014iba}. In this paper, we will address the question of computing 
higher-point null polygons\footnote{I would like to thank Benjamin Basso for informing me about analogous analysis currently under way \cite{BasCaeCorSevVie15} 
following the formalism of Ref.\ \cite{Basso:2014hfa}.} within the OPE framework paying special attention to the factorizability of multiparticle pentagons in terms of single 
particle ones. For identical excitations, its form was conjectured in Ref.\ \cite{Basso:2013vsa} to be 
\begin{align}
\label{FactorizedPentagon}
P (u_1, \dots, u_n | v_1, \dots, v_m)
=
\frac{\prod_{i = 1}^n \prod_{j=1}^m P (u_i|v_j)}{\prod_{i>j}^n P (u_i|u_j) \prod_{k<l}^m P (v_k|v_l)}
\, ,
\end{align}
where the transition is not necessarily particle number-preserving, i.e., $n \neq m$. This was verified for $n=m$ at leading order in Ref.\ \cite{Belitsky:2014rba}
by mapping out the problem of interacting flux-tube excitations to an integrable spin chain with open boundary conditions \cite{Belitsky:2011nn,Sever:2012qp}. Presently, 
the multiparticle pentagons will not be restricted to particles of the same type, however, the factorized form \re{FactorizedPentagon} will still stand strong. 
We will find that the same form is valid for fermions as well where the bootstrap equations are nonlinear. Apart from testing the factorized form, another goal of our
consideration will be to confirm all pentagons introduced in previous analyses. To verify our findings, we will confront them against explicit data on multiparticle 
non-maximally helicity violating (non-MHV) amplitudes. To date, the only available source of the latter\footnote{With the exception of the available symbol for two-loop 
NMHV heptagon derived in Ref.\ \cite{CaronHuot:2011kk}.} is the package by Bourjaily--Caron-Huot--Trnka \cite{Bourjaily:2013mma} that provides amplitudes to 
one-loop order. The main observable will be the ratio function
\begin{align}
\label{RatioFunction}
\mathcal{P}_{N;n} = \mathcal{A}_{N;n}/\mathcal{A}_{N;0}
\, ,
\end{align}
of N$^n$MHV superamplitude $ \mathcal{A}_{N;n}$ to maximally helicity violating (MHV) one $\mathcal{A}_{N;0}$.

Our subsequent presentation will be organized as follows. In the next section, we will focus on the heptagon. We construct a properly subtracted observable from 
the ratio function that is compared with results deduced from OPE. Since, there is one intermediate pentagon in this case, we will address successively single-particle, 
two-to-one and two-to-two transitions in turn. This case alone already encompasses all major flux-tube excitations. Eventually, we take a first look into three-particle 
transitions using a specific example. Next, we analyze octagons following the same footsteps in Sect.\ \ref{OctagonSect}. Finally, we conclude. Several appendices 
are dedicated to provide details of computations performed in the main body. Appendix \ref{AppTwistors} exhibits the construction of reference polygons and the 
way all inequivalent polygons are parametrized. In Appendix \ref{OneLoopPentagonsApp}, we summarize all ingredients of the pentagon approach, i.e., all pentagon
transitions, single and two-particle measures, in the latter case involving one small fermion, as well as flux-tube dispersion relations limiting ourselves to one-loop order.

\section{Heptagon observable}
\label{HeptagonSection}

As we pointed out in the Introduction, the hexagon was exhaustively studied in the literature to a very high order in 't Hooft coupling. Therefore,
to start our present consideration, let us introduce a seven particle observable that we will be comparing our OPE predictions with. For the 
case at  hand, there are two sets of conformal cross ratios $\tau_i$, $\sigma_i$ and $\phi_i$ ($i = 1,2$) which define momentum 
twistors parametrizing inequivalent heptagons \re{HeptagonTwistors}, as discussed in Appendix \ref{AppTwistors}.

While the ratio function $\mathcal{P}_{7;n}$ in Eq.\ \re{RatioFunction} is finite, the OPE frameworks provides predictions for a finite quantity 
which is constructed by factoring out the (inverse) bosonic Wilson loop from the former, namely,
\begin{align}
\label{HeptagonWsubtracted}
\mathcal{W}_{7;n} = g^{2n} \mathcal{P}_{7;n} W_7
\, ,
\end{align}
where, obviously $ \mathcal{P}_{7;0} =1$ and the power of the 't Hooft coupling was introduced to match the definition of the expectation value of the 
supersymmetric Wilson loop of Refs.\ \cite{{Mason:2010yk,CaronHuot:2010ek,Belitsky:2011zm}}, according to which a tree NMHV amplitude corresponds 
to a one-loop expression on the Wilson loop side, N$^2$MHV to two-loop graphs etc. Above, $W_7$ is the heptagon observable that does not depend on 
the Grassmann variables, $W_7 = \mathcal{W}_{7;0}$, and can be split to all orders in 't Hooft coupling as
\begin{align}
W_7 = W_7^{\rm U(1)} \exp (R_7)
\, ,
\end{align}
with $W_7^{\rm U(1)}$ being the sum of connected correlators between reference squares in a chosen tessellation of the heptagon calculated in U(1) 
theory\footnote{See Eq. (127) in Ref.\ \cite{Basso:2013aha}.} with the coupling constant $g^2_{\rm U(1)}$ being replaced by the cusp anomalous dimension 
$g^2_{\rm U(1)} = \ft14 \Gamma_{\rm cusp} (g^2)$ of the full theory \cite{Gaiotto:2011dt} and a remainder function $R_7$. Since in the bulk of the paper we will 
not go beyond the first subleading order in $g^2$ due to the lack of higher loop data for multiparticle amplitudes,  we ignore $R_7$, which starts at 
two loops \cite{DelDuca:2010zg,Goncharov:2010jf}, and use the following expansion that will suffice for our subsequent calculations
\begin{align}
W_7^{\rm U(1)} =1 + g^2 \left[
r_1 (\tau_1, \sigma_1, \phi_1) + r_1 (\tau_2, \sigma_2, \phi_2) + r_2 (\tau_1, \tau_2, \sigma_1, \sigma_2, \phi_1, \phi_2) 
\right]
+ O (g^4)
\, .
\end{align}
The reason for decomposing the order $g^2$ contribution in terms of three functions is that they have a clear representation via single-gluon exchanges 
between reference squares of the abelian Wilson loop \cite{Gaiotto:2011dt}.  At large $\tau$, they develop the following decomposition 
\begin{align}
\label{Wr1}
r_1  (\tau, \sigma, \phi) 
=
{\rm e}^{- \tau} ({\rm e}^{i \phi} + {\rm e}^{-i \phi})
r_{1 [1]} (\sigma)
+ 
{\rm e}^{- 2 \tau} ({\rm e}^{2 i \phi} + {\rm e}^{- 2 i \phi})
r_{1 [2]} (\sigma)
+
\dots
\, .
\end{align}
Here $r_{1 [n]}$ is a twist-$n$ contribution that admits an OPE interpretation in terms of single and two-gluon bound state, respectively,
in the hexagon expansion (or, for the case of the heptagon, as a transition between the flux-tube excitations and the vacuum),
\begin{align}
r_{1 [1]} (\sigma) 
=
\frac{1}{g^2}
\int d \mu_{\rm g} (u) 
\, , \qquad
r_{1 [2]} (\sigma)
=
\frac{1}{g^2}
\int d \mu_{D \rm g} (u)   
\, .
\end{align}
Their perturbative expansion $r = \sum_{\ell \geq 1} g^{2 \ell} r^{(\ell)}$ starts at order $g^2$ and reads\footnote{The emerging here and below one-loop Fourier 
transforms can be computed along the lines of Refs.\ \cite {Papathanasiou:2013uoa,Hatsuda:2014oza}.} at leading order in coupling  \cite{Gaiotto:2011dt,Basso:2014koa}
\begin{align}
r_{1 [1]}^{(1)} (\sigma) 
&= \pi
\int \frac{du}{2 \pi} \frac{-  {\rm e}^{2 i u \sigma}}{( u^2 + \ft14)\cosh (\pi u)}
=
2 \sigma {\rm e}^\sigma - 2 \cosh(\sigma) \ln (1+{\rm e}^{2 \sigma})
\, , \\
r_{1 [2]}^{(1)} (\sigma)
&=
\pi
\int \frac{du}{2 \pi} \frac{u \, {\rm e}^{2 i u \sigma}}{(u^2 + 1) \sinh (\pi u)}
= 
- \frac{1}{2} - \sigma {\rm e}^{2 \sigma} +\cosh(2 \sigma) \ln (1 + {\rm e}^{2 \sigma})
\, ,
\end{align}
making use of the explicit measures summarized in Appendix \ref{OneLoopPentagonsApp}. Analogously, the function $r_2 (\tau_1, \tau_2, \sigma_1, \sigma_2, \phi_1, \phi_2)$ 
that depends on all heptagon variables emerges from the gluon exchange between the top and bottom reference squares and can be expressed in the near-collinear 
limit \cite{Sever:2011pc}
\begin{align}
\label{Wr2}
r_2 (\tau_1, \tau_2, \sigma_1, \sigma_2, \phi_1, \phi_2) 
=
{\rm e}^{- \tau_1 - \tau_2}
({\rm e}^{i \phi_1 + i \phi_2} + {\rm e}^{-i \phi_1 - i \phi_2})
r_{2 [2]} (\sigma_1, \sigma_2)
+ \dots
\, ,
\end{align}
as a gauge field propagating on the flux tube with
\begin{align}
r_{2 [2]} (\sigma_1, \sigma_2)
=
\frac{1}{g^2}
\int d \mu_{\rm g} (u) P_{\rm g|g} (- u| v)  d \mu_{\rm g} (v)
\, ,
\end{align}
involving one intermediate gauge-field pentagon\footnote{The first two terms in its perturbative expansion are given in Appendix \ref{OneLoopPentagonsApp}.} 
$P_{\rm g|g}$ \cite{Basso:2013vsa}. All helicity-violating contributions were not mentioned above since they arise in perturbation theory starting from two-loop order.

Our focus will be on the NMHV amplitudes $\mathcal{P}_{7;1}$. In what follows, we will introduce the following convention for the coefficients in the expansion of the 
heptagon super Wilson loop in a given OPE channel (where we omitted an overall Grassmann structure for each term in the series)
\begin{align}
\mathcal{W}_{7;1} = \sum_{n_1, n_2} \sum_{h_1, h_2} {\rm e}^{- n_1 \tau_1 - n_2 \tau_2} {\rm e}^{ i (h_1 \phi_1 + h_2 \phi_2)/2} 
\mathcal{W}_{[n_1, n_2] (h_1, h_2)} (\sigma_1, \sigma_2; g)
\, .
\end{align}
Here $n_1$ and $n_2$ stand for the cumulative twists of excitations in the incoming and outgoing flux-tube states, while $h_i$ stand for twice their corresponding helicities,
$h_i = 2 m_i$. The function $\mathcal{W}_{[n_1, n_2] (h_1, h_2)} (\sigma_1, \sigma_2; g)$ admits an infinite series expansion in 't Hooft coupling
\begin{align}
\mathcal{W}_{[n_1, n_2] (h_1, h_2)} (\sigma_1, \sigma_2; g)
=
g^2 \sum_{\ell \geq 0} g^{2 \ell} \mathcal{W}^{(\ell)}_{[n_1, n_2] (h_1, h_2)} (\sigma_1, \sigma_2)
\, .
\end{align}
As mentioned earlier, we pulled out a power of the coupling since the one-loop contribution $\mathcal{W}_{7;1}^{(0)}$ to the super Wilson loop is dual to the 
tree-level ratio function $\mathcal{P}_{7;1}^{(0)}$ \cite{CaronHuot:2010ek,Mason:2010yk,Belitsky:2011zm}. Analogously, we will expand the ratio function $\mathcal{P}_{7;1}$,
\begin{align}
\mathcal{P}_{7;1}
= \sum_{n_1, n_2} \sum_{h_1, h_2} {\rm e}^{- n_1 \tau_1 - n_2 \tau_2} {\rm e}^{ i (h_1 \phi_1 + h_2 \phi_2)/2} \mathcal{P}_{[n_1, n_2] (h_1, h_2)} (\sigma_1, \sigma_2; g)
\, ,
\end{align}
with the perturbative series for accompanying coefficients being
\begin{align}
\mathcal{P}_{[n_1, n_2] (h_1, h_2)} (\sigma_1, \sigma_2; g)
=
\sum_{\ell \geq 0} g^{2 \ell} \mathcal{P}^{(\ell)}_{[n_1, n_2] (h_1, h_2)} (\sigma_1, \sigma_2)
\, .
\end{align}

As we alluded to above, the focus of the rest of this section will be on the NMHV contribution $\mathcal{W}_{7;1}$ to the superheptagon. The SU(4) symmetry fixes  
$\mathcal{W}_{7;1}$ to be a homogeneous SU(4) invariant polynomial in Grassmann variables $\chi_i^A$ ($i=1, \dots, 7$) of degree 4. Thus it admits the 
following Grassmann expansion
\begin{align}
\mathcal{W}_{7;1} 
= 
\chi_1^2 \chi_4^2 
\bigg\{
&
{\rm e}^{- \tau_1 - \tau_2}
\mathcal{W}_{[1,1](0,0)} 
\nonumber\\
+\, 
&
{\rm e}^{- 2 \tau_1 - \tau_2} 
{\rm e}^{i \phi_1} 
\mathcal{W}_{[2,1](2,0)} 
+
{\rm e}^{- 2 \tau_1 - \tau_2} 
{\rm e}^{- i \phi_1} 
\mathcal{W}_{[2,1](- 2,0)} 
\nonumber\\
&\qquad\qquad\qquad\qquad\ \ \,
+
{\rm e}^{- 2 \tau_1 - 2 \tau_2} 
\left[
{\rm e}^{- i \phi_1 + i \phi_2} 
\mathcal{W}_{[2,2](- 2,2)} 
+
\dots
\right]
+ \dots
\bigg\}
\nonumber\\
+
\chi_1^3 \chi_4 
\bigg\{
&
{\rm e}^{- \tau_1 - \tau_2} {\rm e}^{i \phi_1/2 + i \phi_2/2} 
\mathcal{W}_{[1,1](1,1)} 
\nonumber\\
+ \, 
& 
{\rm e}^{- 2\tau_1 - \tau_2} 
\left[
{\rm e}^{3 i \phi_1/2 + i \phi_2/2} 
\mathcal{W}_{[2,1](3,1)} 
+
{\rm e}^{- i \phi_1/2 + i \phi_2/2} 
\mathcal{W}_{[2,1](-1,1)} 
\right]
\nonumber\\ 
+ \, 
& 
{\rm e}^{- 2 \tau_1 - 2 \tau_2} 
\big[
{\rm e}^{3 i \phi_1/2 + 3 i \phi_2/2} 
\mathcal{W}_{[2,2](3,3)} 
+
{\rm e}^{- i \phi_1/2 + 3 i \phi_2/2} 
\mathcal{W}_{[2,2](-1,3)} 
+
\dots
\big]
\nonumber\\
+ \, 
& 
{\rm e}^{- 3 \tau_1 - \tau_2} 
\left[
{\rm e}^{- 3 i \phi_1/2 + i \phi_2/2} 
\mathcal{W}_{[3,1](- 3,1)} 
+
\dots
\right]
+ \dots
\bigg\} 
\nonumber\\
+ \dots \, , \quad&
\end{align}
where $\chi_1^2 \chi_4^2 = \varepsilon_{ABCD} \chi_1^A \chi_1^B \chi_4^C \chi_4^D$ etc. Here we only displayed terms in the OPE that form a representative 
class of contributions that will be the subject of the current analysis. The $\chi^4_i$ contributions were considered previously in Ref.\ \cite{Basso:2014nra} and 
thus to avoid repetition will not be discussed below.

\subsection{One-to-one transitions}

We start our consideration with the $\chi_1^2 \chi_4^2$ component. The leading twist contribution is determined by the exchange of the hole excitation propagating 
on the color flux tube and, as was established in Ref.\ \cite{Basso:2013aha}, it reads
\begin{align}
\mathcal{W}_{[1,1](0,0)}
=
-
\int d \mu_{\rm h} (u) P_{\rm h|h} (-u|v) d \mu_{\rm h} (v)
\, .
\end{align}
To lowest order in perturbative series, it takes the form
\begin{align}
\mathcal{W}^{(0)}_{[1,1](0,0)} (\sigma_1, \sigma_2) 
=
\mathcal{P}^{(0)}_{[1,1](0,0)} (\sigma_1, \sigma_2) 
&
=
-
\pi^2 \int \frac{d u}{2 \pi} \frac{d v}{2 \pi} \frac{{\rm e}^{2 i \sigma_1 u + 2 i \sigma_2 v}}{\cosh (\pi u) \cosh (\pi v)}
\frac{\Gamma (- i u - i v)}{\Gamma (\ft12 - i u) \Gamma (\ft12 - i v)}
\nonumber\\
&
=  
-
\frac{{\rm e}^{\sigma_1 + \sigma_2}}{{\rm e}^{2 \sigma_1} +  {\rm e}^{2 \sigma_2} + {\rm e}^{2 \sigma_1 + 2 \sigma_2}}
\, ,
\end{align}
making use of results summarized in Appendix \ref{OneLoopPentagonsApp}. It agrees with the expression for the ratio function $\mathcal{P}^{(0)}_{[1,1](0,0)}$
deduced from the package \cite{Bourjaily:2013mma}. The subleading correction in 't Hooft coupling agrees as well. We do not display it explicitly here in order
to save space (see, however, the ancillary file).

Next, we turn to the $\chi_1^3 \chi_4$ component of the heptagon. As can be easily anticipated on the basis of quantum numbers by counting the 
fermionic degree and SU(4) weight of the accompanying Grassmann structure, the leading term in the near-collinear expansion is governed by the exchange 
of the fermionic flux-tube excitation, namely,
\begin{align}
\mathcal{W}_{[1,1](1,1)}
= - i
\int d \mu_{\Psi} (u_1) x[u_1] P_{\Psi|\Psi} (-u_1|v_1) \, d \mu_\Psi (v_1)
\, ,
\end{align}
where $x[u]$ is an ad hoc NMHV fermionic form factor \cite{Basso:2014koa,Belitsky:2014sla,Belitsky:2014lta} given by the Zhukowski variable whose definition is
deferred to Eq.\ \re{ZhukowskiVariable} of Appendix \ref{OneLoopPentagonsApp}. Here the contour $C$ runs 
on both the large and small fermion sheets of the corresponding Riemann surface, as was reviewed in the Introduction. Since there are no poles in the integrand on the 
small fermion sheet, its contribution vanishes identically. This can be understood recalling that the small fermion at zero momentum is a 
generator of supersymmeric transformation \cite{Alday:2007mf}. Since for the case at hand, it is the only excitation present on the top or the bottom of the heptagon, it 
would correspond to the action of the supersymmetry generator on the vacuum state and thus yield vanishing net result. Hence, the above super Wilson loop component 
reads in the OPE framework
\begin{align}
\mathcal{W}_{[1,1](1,1)}
=
- i
\int d \mu_{\rm F} (u) x[u] P_{\rm F|F} (-u|v) d \mu_{\rm F} (v)
\, .
\end{align}
Taylor expanding all ingredients in 't Hooft coupling, we immediately reproduce the tree-level $\chi_1^3 \chi_4$ contribution to the ratio function
\begin{align}
\mathcal{W}^{(0)}_{[1,1](1,1)} (\sigma_1, \sigma_2) 
=
\mathcal{P}^{(0)}_{[1,1](1,1)} (\sigma_1, \sigma_2) 
&
=
- \pi^2 \int \frac{d u_1}{2 \pi} \frac{d v_1}{2 \pi} \frac{{\rm e}^{2 i \sigma_1 u_1 + 2 i \sigma_2 v_1}}{\sinh (\pi u_1) \sinh (\pi v_1)}
\frac{\Gamma (- i u_1 - i v_1)}{\Gamma (- i u_1) \Gamma (1 - i v_1)}
\nonumber\\
&
= \frac{{\rm e}^{2 \sigma_1}}{(1 + {\rm e}^{2 \sigma_1}) ({\rm e}^{2 \sigma_1} +  {\rm e}^{2 \sigma_2} + {\rm e}^{2 \sigma_1 + 2 \sigma_2})}
\, .
\end{align}
Substituting perturbative expansions quoted in Appendix \ref{OneLoopPentagonsApp}, this agreement can be extended to one loop order as well (see the accompanying
Mathematica notebook).

Analogously, if we were to consider the $\chi_1 \chi_4^3$ component, one would find that the leading twist contribution is determined by the permutation transformation of 
the $\chi_1^3 \chi_4$ contribution, i.e., $\mathcal{W}_{[1,1](- 1,- 1)} (\sigma_1, \sigma_2; g)  = \mathcal{W}_{[1,1](1,1)} (\sigma_2, \sigma_1; g) $ and reads explicitly,
\begin{align}
\mathcal{W}_{[1,1](- 1,- 1)}
=
- i
\int d \mu_{\rm F} (u)P_{\rm F|F} (-u|v) x[v] d \mu_{\rm F} (v)
\, .
\end{align}

\subsection{Two-to-one transitions}

Let us move on to two-particle states. We begin our discussion of twist-two contributions with the particle number-changing case when the bottom part of the 
heptagon emits two flux-tube excitations while the top absorbs only one. We will observe on a number of examples that the two-to-one pentagons factorize in terms of
single-particle ones as follows\footnote{We do not display potential kinematical SU(4) tensor structure but focus only on the dynamical part of pentagon transitions.}
\begin{align}
\label{2to1Pentagon}
P_{\rm p_1p_2|p_3} (u_1, u_2| v_1) 
=
\frac{P_{\rm p_1 | p_3} (u_1| v_1)  P_{\rm p_2 | p_3} (u_2| v_1) }{P_{\rm p_2 | p_1} (u_2 | u_1) }
\, . 
\end{align}
The above expression follows the pattern of gluonic transitions conjectured in Ref.\ \cite{Basso:2013vsa} and verified at leading order in Ref.\ \cite{Belitsky:2014rba}, 
however, here it is not restricted to particles of the same type. We will find that the same form is valid for fermions where the bootstrap equations are nonlinear 
\cite{Basso:2014koa,Belitsky:2014lta}.

\subsubsection{Two-(anti)fermion and scalar-(anti)gluon states}
\label{Tw2Tw1PsiBarPsiBarSection}

To analyze the two-(anti)fermion and scalar-(anti)gluon states, we turn to the twist-two contribution in the $\chi_1^2 \chi_4^2$ Grassmann component.
The operator product expansion for $\mathcal{W}_{[2,1] (2,0)}$ is related to the properly defined ratio function \re{HeptagonWsubtracted} as follows
\begin{align}
\label{RatFuncW12120tree}
\mathcal{W}^{(0)}_{[2,1] (2,0)} (\sigma_1, \sigma_2)
&= \mathcal{P}^{(0)}_{[2,1] (2,0)} (\sigma_1, \sigma_2)
\, ,\\
\label{RatFuncW12120}
\mathcal{W}^{(1)}_{[2,1] (2,0)}  (\sigma_1, \sigma_2)
&= \mathcal{P}^{(1)}_{[2,1] (2,0)} (\sigma_1, \sigma_2) 
+ 
\mathcal{P}^{(0)}_{[1,1] (0,0)} (\sigma_1, \sigma_2) r_{1 [1]}^{(1)} (\sigma_1) 
\, ,
\end{align}
at tree and one-loop order, respectively. These arise from the sum of two particles that mimic the quantum numbers of a hole in the in-state and a single hole in 
the outgoing state, such that
\begin{align}
\mathcal{W}_{[2,1](2,0)} = \mathcal{W}_{\Psi\Psi|{\rm h}} + \mathcal{W}_{\rm gh|h}
\, , 
\end{align}
with
\begin{align}
\mathcal{W}_{\Psi\Psi|{\rm h}} 
&
= \int
d \mu_{\Psi} (u_1) d \mu_{\Psi} (u_2) \frac{[ x [u_1] x [u_2] ]^{3/2}}{g^4} F^{\bf 6}_{\Psi\Psi} (0 | u_1, u_2)
P_{\Psi\Psi| \rm h} (- u_2, - u_1| v_1) d \mu_{\rm h} (v_1)
\, , \\
\mathcal{W}_{\rm gh|h}
&=
\int d \mu_{\rm g} (u_1) d \mu_{\rm h} (u_2) \, \frac{\sqrt{x^+[u_1] x^-[u_1]}}{g} F_{\rm gh} (0|u_1, u_2) P_{\rm hg| h} (- u_2, -u_1| v_1) d \mu_{\rm h} (v_1)
\, , 
\end{align}
where the two-particle production form factors are \cite{Belitsky:2014sla,Belitsky:2014lta}
\begin{align}
\label{PsiPsiFF} 
F^{\bf 6}_{\Psi\Psi} (0 | u_1, u_2)
=
\frac{i}{u_1 - u_2 + i} \frac{1}{P_{\Psi|\Psi} (u_1| u_2)}
\, , \qquad
F_{\rm gh} (0 | u_1, u_2)
=
\frac{i}{P_{\rm g| h} (u_1| u_2)}
\, .
\end{align}
In the course of the study, we established the following empirical rule for introduction of additional ``helicity'' form factors, whenever the intermediate pentagon 
transitions $P_{\rm p_1, p_2, \dots | p^\prime_1, p^\prime_2, \dots}$ involved (anti)gluons, ${\rm p}_i, {\rm p}^\prime_j = {\rm g}, \bar{\rm g}$. Namely, for each 
pair in the product of all permutations $\sigma = \{1,2,\dots\}$ and $\sigma^\prime = \{1^\prime, 2^\prime, \dots\}$ of in-out state transitions ${\rm p}_{\sigma} | 
{\rm p}^\prime_{\sigma^\prime}$ , we introduced extra factors depending on the shifted Zhukowski variables \re{ShiftedZhukowski} according to the rule
\begin{align}
\label{PsiGannihilationFF}
\bra{{\rm g} (u)}\, , \ket{\bar{\rm g} (u)} \to  \frac{g}{\sqrt{x^+[u] x^- [u]}}
\, , \qquad
\bra{\bar{\rm g} (u)}\, , \ket{{\rm g} (u)} \to  \frac{\sqrt{x^+[u] x^- [u]}}{g}
\, .
\end{align}
when the in/out state contained an (anti)gluon, and the conjugate one had a hole and an (anti)fermion. This compensated the square-root factors emerging in the solution 
to bootstrap equations in the conventions of Refs.\ \cite{Belitsky:2014sla,Belitsky:2014lta}.

In the same fashion as in the previously addressed twist-one case, the fermionic contour runs on both sheets of the Riemann surface. Presently, the leading 
effect arises, however, from the kinematics when one of the rapidities belongs to the small sheet and another to the large one. Since $ F^{\bf 6}_{\rm fF} (0 | u_1, u_2)$ 
has a pole in the lower half plane at $u_1 = u_2 - i$, one can evaluate the integral over $u_1$ using the Cauchy theorem. Thus $\mathcal{W}_{\Psi\Psi|{\rm h}}$ splits
up into the sum of two contributions, $\mathcal{W}_{\Psi\Psi|{\rm h}} = \mathcal{W}_{\rm fF|h} + \mathcal{W}_{\rm FF|h}$, 
\begin{align}
\label{fFhW}
\mathcal{W}_{\rm fF|h}
&
= 
\frac{1}{g^2}
\int
d \mu_{\rm fF} (u_2) \sqrt{ x[u_2]/x[u_2 - i] }
P_{\rm f|h} (- u_2 + i| v_1)P_{\rm F|h} (- u_2 | v_1) d \mu_{\rm h} (v_1)
\, , \\
\mathcal{W}_{\rm FF|h}
&
= 
\frac{1}{g^4}
\int
d \mu_{\rm F} (u_1) d \mu_{\rm F} (u_2)
\frac{i  [ x [u_1] x [u_2] ]^{3/2} P_{\rm F|h} (- u_1| v_1) P_{\rm F|h} (- u_2| v_1)}{(u_1 - u_2 + i) P_{\rm F|F} (u_1|u_2)  P_{\rm F|F} (- u_1|- u_2)} d \mu_{\rm h} (v_1)
\, , 
\end{align}
of the small-large and large-large fermion pairs, respectively. In the former, we employed a notation for the composite two-fermion measure  \cite{Belitsky:2014sla}
\begin{align}
\label{mufF}
\mu_{\rm fF} (u) 
=
- 
\frac{x[u]}{x[u-i]} \frac{\mu_{\rm f} (u-i) \mu_{\rm F} (u)}{P_{\rm f|F} (u-i|u) P_{\rm f|F} (- u+i|-u)}
\, .
\end{align}
It is important to recall that as one passes to the small fermion sheet, the Zhukowski variable $x$ transforms to $g^2/x$ \cite{Basso:2010in}. Equation \re{fFhW} accounts 
for the entire tree-level NMHV ratio function as it starts at order $g^2$, 
\begin{align}
\mathcal{W}^{(0)}_{[2,1] (2,0)} (\sigma_1, \sigma_2)
=
\mathcal{W}^{(0)}_{\rm fF|h}  (\sigma_1, \sigma_2)
&
=
\pi^2
\int \frac{d u_2}{2 \pi} \frac{d v_1}{2 \pi} \frac{{\rm e}^{2 i u_2 \sigma_1 + 2 i v_1 \sigma_2}}{\sinh (\pi u_2) \cosh (\pi v_1)}
\frac{i \Gamma (\ft12 -  i u_2 - i v_1)}{\Gamma (- i u_2) \Gamma (\ft12 -  i v_1)}
\nonumber\\
&
=
\frac{{\rm e}^{2 \sigma_1 + 3 \sigma_2}}{({\rm e}^{2 \sigma_1} + {\rm e}^{2 \sigma_2} + {\rm e}^{2 \sigma_1 + 2 \sigma_2})^2}
\, .
\end{align}
While the second one, $\mathcal{W}_{\rm FF|h}$, is postponed to $O (g^4)$. Notice that the latter contributes to the Wilson loop an order earlier 
in 't Hooft coupling compared to the hexagon \cite{Belitsky:2014sla}. In addition, the one-loop ratio function \re{RatFuncW12120} receives an 
additive contribution from the incoming gluon-hole state,
\begin{align}
\mathcal{W}_{\rm gh|h}
&=
\frac{1}{g}
\int d \mu_{\rm g} (u_1) d \mu_{\rm h} (u_2) \frac{i \sqrt{x^+[u_1] x^-[u_1]}
P_{\rm g| h} (-u_1| v_1) P_{\rm h| h} (-u_2| v_1)}{P_{\rm g| h} (u_1| u_2) P_{\rm g| h} (- u_1| - u_2)} d \mu_{\rm h} (v_1)
\, .
\end{align}
Combining $\mathcal{W}^{(1)}_{\rm fF|h}$, $\mathcal{W}^{(1)}_{\rm FF|h}$ and $\mathcal{W}^{(1)}_{\rm gh|h}$ together, we uncover the one-loop NMHV amplitude 
\re{RatFuncW12120} as demonstrated in the accompanying Mathematica notebook.

The two-antifermion states emerge in the $\mathcal{W}_{[2,1] (- 2,0)}$ component of the superloop. This transition will be sensitive to the hole-antifermion
pentagons. Since the bosonic Wilson loop is symmetric with respect to the flip in sign of the gluon helicity, i.e., $\phi_1 \leftrightarrow - \phi_1$, the subtracted ratio 
function takes the same form as above Eqs.\ \re{RatFuncW12120tree} and \re{RatFuncW12120} with obvious substitutions $\mathcal{W}^{(\ell)}_{[2,1] (2,0)} \to 
\mathcal{W}^{(\ell)}_{[2,1] (- 2,0)}$. These arise from the sum of two-particles in the in-state and a single hole in the outgoing state,
such that
\begin{align}
\mathcal{W}_{[2,1](- 2,0)} = \mathcal{W}_{\bar\Psi\bar\Psi|{\rm h}} + \mathcal{W}_{\rm \bar{g}h|h}
\, , 
\end{align}
with
\begin{align}
\mathcal{W}_{\bar\Psi\bar\Psi|{\rm h}} 
&
= 
\int
d \mu_{\Psi} (u_1) d \mu_{\Psi} (u_2) 
\frac{\sqrt{x[u_1] x[u_2]}}{g^2}
F^{\bf 6}_{\Psi\Psi} (0 | u_1, u_2)
P_{\bar\Psi\bar\Psi| \rm h} (- u_2, - u_1| v_1) d \mu_{\rm h} (v_1)
\, , \\
\mathcal{W}_{\rm \bar{g}h|h}
&=
\int d \mu_{\rm g} (u_1) d \mu_{\rm h} (u_2) \frac{g F_{\rm \bar{g}h} (0|u_1, u_2)}{\sqrt{x^+ [u_1] x^- [u_1]}} P_{\rm h\bar{g}| h} (- u_2, -u_1| v_1) d \mu_{\rm h} (v_1)
\, .
\end{align}
However, an immediate inspection demonstrates that to one-loop accuracy, the small-large antifermion pair alone accommodates the entire
contribution in the operator product expansion such that to this accuracy $\mathcal{W}_{[2,1](- 2,0)}$ reads
\begin{align}
\label{fbarFbarhW}
\mathcal{W}_{[2,1](- 2,0)} 
&
=
\mathcal{W}_{\rm \bar{f}\bar{F}|h}
+ O(g^6)
\\
&
= 
\frac{1}{g^2}
\int
d \mu_{\rm fF} (u_2) \sqrt{x[u_2 - i]/x[u_2]} P_{\rm f|h} (- u_2 + i| v_1)P_{\rm F|h} (- u_2 | v_1) d \mu_{\rm h} (v_1)
+ O(g^6)
\, . \nonumber
\end{align}
Perturbative expansion yields for the tree amplitude
\begin{align}
\mathcal{W}^{(0)}_{[2,1](- 2,0)} 
=
\mathcal{W}^{(0)}_{\rm \bar{f}\bar{F}|h}
&
=
\pi^2 \int \frac{d u_2}{2 \pi} \frac{d v_1}{2 \pi} \frac{{\rm e}^{2 i u_2 \sigma_1 + 2 i v_1 \sigma_2}}{\sinh (\pi u_2) \cosh (\pi v_1)}
\frac{(u_2 - i) \Gamma (\ft12 -  i u_2 - i v_1)}{\Gamma (1 - i u_2) \Gamma (\ft12 -  i v_1)}
\nonumber\\
&
=
- \frac{{\rm e}^{5 \sigma_2}}{(1+ {\rm e}^{2 \sigma_2})({\rm e}^{2 \sigma_1} + {\rm e}^{2 \sigma_2} + {\rm e}^{2 \sigma_1 + 2 \sigma_2})^2}
\, . 
\end{align}
Further expansion in 't Hooft coupling making use of explicit integrability input from Appendix \ref{OneLoopPentagonsApp} shows that this 
small-large antifermion pair solely accounts for $\mathcal{W}^{(1)}_{[2,1](- 2,0)}$ as well. The true two-particle contribution get pushed to two-loop
order similarly to the phenomenon observed for the NMHV hexagon \cite{Belitsky:2014sla,Belitsky:2014lta}.

\subsubsection{Antifermion-hole states}
\label{AntiFermionHoleHeptagonSection}

The above consideration exhausted the two-to-one transitions to the $\chi_1^2 \chi_4^2$ Grassmann component, thus we turn without further ado
to the $\chi_1^3 \chi_4$ contribution. The quantum numbers of states propagating in this OPE channel suggest that $\mathcal{W}_{[2,1](-1,1)}$ is 
determined by the sum of an antifermion-hole and antigluon-fermion produced at the bottom of the Wilson loop,
\begin{align}
\mathcal{W}_{[2,1](-1,1)}
=
\mathcal{W}_{\bar{\Psi}{\rm h}| \Psi}
+
\mathcal{W}_{\Psi \bar{\rm g}| \Psi}
\, .
\end{align}
Their explicit all-order expression is given by
\begin{align}
\mathcal{W}_{\bar{\Psi}{\rm h}| \Psi}
&
=
\int d \mu_\Psi (u_1)\, d \mu_{\rm h} (u_2) \, F^{\bf 4}_{\bar\Psi {\rm h}} (0| u_1, u_2)
P_{{\rm h} \bar\Psi | \Psi} (- u_2, - u_1 | v_1)  \, \frac{i \sqrt{x[v_1]}}{g} d \mu_\Psi (v_1)
\, , \\
\mathcal{W}_{\Psi \bar{\rm g}| \Psi}
&
=
\int d \mu_\Psi (u_1)\, d \mu_{\rm g} (u_2) \, i x[u_1] \frac{g F_{\Psi \bar{\rm g}} (0| u_1, u_2)}{\sqrt{x^+ [u_2] x^- [u_1]}}
P_{\bar{\rm g} \Psi | \Psi} (- u_2, - u_1 | v_1)  \, d \mu_\Psi (v_1)
\, .
\end{align}
Here an (anti)gluon accompanies a flux-tube fermion as a consequence, we introduced an additional form factor according to the rule \re{PsiGannihilationFF}.
In our previous studies \cite{Belitsky:2014lta}, we found that the form factor $F^{\bf 4}_{\bar\Psi {\rm h}} (0| u_1, u_2)$ possesses a pole in the lower half-plane 
of the small fermion Riemann sheet, while the $F_{\Psi \bar{\rm g}} (0| u_1, u_2)$ one does not. This implies that $\mathcal{W}_{\bar{\Psi}{\rm h}| \Psi}$ at least 
induces the entire tree-level ratio function $\mathcal{P}^{(0)}_{[2,1](-1,1)}$. Substituting
\begin{align}
F^{\bf 4}_{\bar\Psi {\rm h}} (0| u_1, u_2) = \frac{1}{u_1 - u_2 + \ft{3 i}{2}} \frac{1}{P_{\bar\Psi|{\rm h}} (u_1 | u_2)}
\, ,
\end{align}
and splitting $\mathcal{W}_{\bar{\Psi}{\rm h}| \Psi}$ into the sum of the small and large fermions in the initial state, $\mathcal{W}_{\bar{\Psi}{\rm h}| \Psi}
= \mathcal{W}_{\bar{\rm f}{\rm h}| {\rm F}} + \mathcal{W}_{\bar{\rm F}{\rm h}| {\rm F}}$, we find that $\mathcal{W}_{\bar{\rm f}{\rm h}| {\rm F}}$, that reads
\begin{align}
\mathcal{W}_{\bar{\rm f}{\rm h}| {\rm F}}
=
- \frac{1}{g}
\int
d \mu_{\rm fh} (u_2) P_{\rm \bar{f}|F} (- u_2 + \ft{3 i}{2}| v_1)P_{\rm h|F} (- u_2 | v_1)  \sqrt{x[v_1]} d \mu_{\rm F} (v_1)
\, ,
\end{align}
actually does account for both tree and one-loop subtracted ratio function \re{HeptagonWsubtracted}. Here we used a notation for the composite small-antifermion--hole 
measure \cite{Belitsky:2014lta}
\begin{align}
\mu_{\rm hf} (u) = \frac{g^2 \mu_{\rm h} (u) \mu_{\rm f} (u - \ft{3 i}{2})}{P_{\rm h|f} \left(u|u - \ft{3i}{2}\right) P_{\rm h|f} \left(-u|-u + \ft{3i}{2} \right)}
\, ,
\end{align}
Employing its perturbative expansion summarized in Appendix \ref{OneLoopPentagonsApp}, we find
\begin{align}
\mathcal{W}^{(0)}_{[2,1](-1,1)}
=
\mathcal{W}^{(0)}_{\bar{\rm f}{\rm h}| {\rm F}}
&
=
\pi^2 \int \frac{d u_2}{2 \pi} \frac{d v_1}{2 \pi} \frac{{\rm e}^{2 i u_2 \sigma_1 + 2 i v_1 \sigma_2}}{\cosh (\pi u_2) \sinh (\pi v_1)}
\frac{(u_2 - \ft{3 i}{2})\Gamma (\ft12 -  i u_2 - i v_1)}{\Gamma (\ft12 - i u_2) \Gamma (1 -  i v_1)}
\nonumber\\
&
=
- \frac{
{\rm e}^{3 \sigma_1} 
(
2 {\rm e}^{2 \sigma_1} + {\rm e}^{4 \sigma_1}  + 3 {\rm e}^{2 \sigma_2} + 4 {\rm e}^{2 \sigma_1 + 2 \sigma_2} + {\rm e}^{4 \sigma_1 + 2 \sigma_2}
)
}{
(1 + {\rm e}^{2 \sigma_1})^2
({\rm e}^{2 \sigma_1} + {\rm e}^{2 \sigma_2} + {\rm e}^{2 \sigma_1 + 2 \sigma_2})^2
}
\, .
\end{align}
We also observed agreement for $\mathcal{W}^{(1)}_{[2,1](-1,1)}$ at one loop order (see the ancillary file). This implies that contributions from the 
large-antifermion-hole and antigluon-fermion pairs cumulatively vanish at this order.

\subsubsection{Fermion-gluon states}

We continue the analysis of the $\chi_1^3 \chi_4$ component by unravelling the structure of its $\mathcal{W}_{[2,1](3,1)}$ term. The tree and one-loop 
coefficients in the perturbative expansion of the latter are related to the ratio function via Eqs.\ \re{RatFuncW12120tree} and \re{RatFuncW12120} 
with $\mathcal{W}^{(\ell)}_{[2,1] (2,0)}$ replaced by $\mathcal{W}^{(\ell)}_{[2,1] (3,1)}$ and $\mathcal{P}^{(0)}_{[1,1] (0,0)}$ by $\mathcal{P}^{(0)}_{[1,1] (1,1)}$.
The analysis of quantum numbers suggests that this component is induced by the emission of the gluon-fermion pair at the bottom of the hexagon 
and absorption of a single fermion at the top. Thus its all-order expression reads
\begin{align}
\mathcal{W}_{[2,1](3,1)} 
=
\mathcal{W}_{\Psi{\rm g}|\Psi}
=
\int d \mu_\Psi (u_1) \, d \mu_{\rm g} (u_2) & \, i x[u_1] \, \frac{\sqrt{x^+ [u_2] x^- [u_2]}}{g}
\\
&\times F_{\Psi {\rm g}} (0| u_1, u_2)
P_{{\rm g} \Psi | \Psi} (- u_2, - u_1 | v_1) \, d \mu_\Psi (v_1)
\, . \nonumber
\end{align}
Here the form factor for the production of the $\Psi {\rm g}$ state is
\begin{align}
\label{PsiGproductionFF}
F_{\Psi {\rm g}} (0| u_1, u_2)
=
\frac{1}{P_{\Psi |{\rm g}} (u_1| u_2)}
\, .
\end{align}
The fermion contour runs over the large and small fermion sheets such that 
\begin{align}
\mathcal{W}_{[2,1](3,1)} = \mathcal{W}_{{\rm fg}| {\rm F}} + \mathcal{W}_{{\rm Fg}| {\rm F}}
\, .
\end{align}
Due to a zero in the small-fermion pentagon $P_{{\rm f} |{\rm g}} (u_1| u_2)$, the corresponding form factor possesses a pole at 
$u_1 = u_2 - \ft{i}{2}$ in the lower half plane of the small fermion Riemann sheet. Evaluating the integral over the small fermion rapidity 
$u_1$, we obtain the expression
\begin{align}
\mathcal{W}_{{\rm fg}| {\rm F}}
=
\frac{1}{g}
\int
d \mu_{\rm fg} (u_2) \sqrt{x^+ [u_2] x^- [u_2]} P_{\rm f|F} (- u_2 + \ft{i}{2}| v_1)P_{\rm g|F} (- u_2 | v_1) d \mu_{\rm F} (v_1)
\, , 
\end{align}
where we introduced, following Ref.\ \cite{Belitsky:2014lta}, the composite fermion-gluon measure
\begin{align}
\mu_{\rm fg} (u)
=
i g^2 \frac{ \mu_{\rm g} (u) \mu_{\rm f} (u^-)}{x[u^-]}\frac{\bar{f}_{\rm \bar{g}f} (u, u^-) \bar{f}_{\rm \bar{g}f} (-u,-u^-)}{P_{\rm \bar{g}|f} (u|u^-) P_{\rm \bar{g}|f} (-u|-u^-)}
\, ,
\end{align}
with $\bar{f}_{\rm \bar{g}f} (u, v) = (x^+[u] x [v] - g^2) (x^- [u] x[v] - g^2)/(g^2 x[v])$.
Making use of their Taylor expansion in coupling, one can easily convince oneself that the tree and one-loop terms of $\mathcal{W}_{{\rm fg}| {\rm F}}$ 
reproduce the subtracted ratio function $\mathcal{W}_{[2,1](3,1)}$ with the leading term yielding explicitly
\begin{align}
\mathcal{W}^{(0)}_{[2,1](3,1)}
&
=
\pi^2 \int \frac{d u_2}{2 \pi} \frac{d v_1}{2 \pi} \frac{{\rm e}^{2 i u_2 \sigma_1 + 2 i v_1 \sigma_2}}{\cosh (\pi u_2) \sinh (\pi v_1)}
\frac{i \Gamma (\ft12 -  i u_2 - i v_1)}{\Gamma (- \ft12 - i u_2) \Gamma (1 -  i v_1)}
\nonumber\\
&
=
- \frac{{\rm e}^{3 \sigma_1} ({\rm e}^{2 \sigma_1} + 2 {\rm e}^{2 \sigma_2} + 2 {\rm e}^{2 \sigma_1 + 2 \sigma_2})
}{
(1 + {\rm e}^{2 \sigma_1})^2
({\rm e}^{2 \sigma_1} + {\rm e}^{2 \sigma_2} + {\rm e}^{2 \sigma_1 + 2 \sigma_2})^2
}
\, .
\end{align}
The contribution of the large fermion--gluon pair is postponed to two-loop order as verified in the accompanying notebook. The twist-one-to-twist-two 
transition in the $\chi_1 \chi_4^3$ component, i.e., ${\rm e}^{- \tau_1 - 2 \tau_2} {\rm e}^{- i \phi_1/2 - 3 i \phi_2/2}$ is eagerly obtained from the above 
by the $\sigma_1 \leftrightarrow \sigma_2$ interchange.

\subsection{Two-to-two transitions}

To elucidate the factorizability of multiparticle pentagons even further, let us address a couple of examples involving two-to-two transitions. For these, the 
dynamical part of multiparticle pentagons will be assumed to admit the following form, echoing Eq.\ \re{FactorizedPentagon},
\begin{align}
\label{2to2Pentagon}
P_{\rm p_1p_2|p_3 p_4} (u_1, u_2| v_1, v_2) 
=
\frac{P_{\rm p_1 | p_3} (u_1| v_1)  P_{\rm p_2 | p_3} (u_2| v_1) P_{\rm p_1 | p_4} (u_1| v_2) P_{\rm p_2 | p_4} (u_2| v_2) }{P_{\rm p_2 | p_1} (u_2 | u_1) P_{\rm p_3 | p_4} (v_1 | v_2) }
\, . 
\end{align}
Let us offer a perturbative confirmation for this conjecture though a number of examples.

\subsubsection{Two-(anti)fermion states}

First, we will analyze the $\mathcal{W}_{[2,2](- 2,2)}$ term in the $\chi_1^2 \chi_4^2$ component of the superloop. The function $\mathcal{W}_{[2,2](- 2,2)}$ can be split 
into the sum of four terms
\begin{align}
\mathcal{W}_{[2,2](- 2,2)}
=
\mathcal{W}_{\bar\Psi\bar\Psi|\Psi\Psi} + \mathcal{W}_{{\rm \bar{g}h}|\Psi\Psi} + \mathcal{W}_{\bar\Psi\bar\Psi|{\rm g h}} + \mathcal{W}_{{\rm \bar{g}h}|{\rm gh}} 
\, .
\end{align}
However, only the first term in the right-hand side induces a nontrivial contribution to the first two orders in the perturbative expansion of the ratio function as will be 
established momentarily. This two-antifermion-to-two-fermion transition admits the following form in terms of flux-tube pentagons
\begin{align}
\mathcal{W}_{\bar\Psi\bar\Psi|\Psi\Psi}
=
\int 
&
d \mu_{\Psi} (u_1) d \mu_{\Psi} (u_2) 
\frac{x[u_1]}{g^2}\frac{x[u_2]}{g^2}
F^{\bf 6}_{\Psi\Psi} (0 | u_1, u_2)
\\
&
\times
P_{\bar\Psi\bar\Psi|\Psi\Psi} (-u_2, - u_1 | v_1, v_2)
\frac{x[v_1]}{g^2}\frac{x[v_2]}{g^2}
F^{\bf 6}_{\Psi\Psi} (- v_2, - v_1|0)
\mu_{\Psi} (v_1) d \mu_{\Psi} (v_2)
\nonumber
\end{align}
and can be split into four terms depending on whether the fermion rapidity belongs to the large or small fermion sheet. As a working hypothesis, the two-to-two particle 
pentagon will be taken in the factorized form \re{2to2Pentagon}. In the above formula, the absorption form factor is related to the emission one \re{PsiPsiFF} via 
$F^{\bf 6}_{\Psi\Psi} (- v_2, - v_1|0) = F^{\bf 6}_{\Psi\Psi} (0|v_2, v_1)$. Since the two-fermion production/absorption form factors possess poles and the rest of the 
integrand is a holomorphic function, the integrals over the small fermion $u_1$ and $v_2$ rapidities can be worked out by calculating the residues at the position 
of the latter yielding
\begin{align}
\mathcal{W}_{\rm \bar{f}\bar{F}|Ff}
=
\frac{1}{g^2}
\int d \mu_{\rm fF} (u_2)
P_{\rm \bar{F}|F} (-u_2|v_1)
P_{\rm \bar{F}|f} (-u_2|v_1 - i)
P_{\rm \bar{f}|F} (-u_2 + i|v_1)
P_{\rm \bar{f}|f} (-u_2 + i|v_1 - i)
d \mu_{\rm fF} (v_1)
\, .
\end{align}
Here the composite small-large fermion measure was quoted earlier in Eq.\ \re{mufF}. Making use of the perturbative expansion summarized in Appendix 
\ref{OneLoopPentagonsApp}, we find a complete agreement with subtracted tree and one-loop ratio functions
\begin{align}
\mathcal{W}^{(0)}_{[2,2] (-2,2)} (\sigma_1, \sigma_2)
&= \mathcal{P}^{(0)}_{[2,2] (-2,2)} (\sigma_1, \sigma_2)
\, ,\\
\mathcal{W}^{(1)}_{[2,2] (-2,2)}  (\sigma_1, \sigma_2)
&= \mathcal{P}^{(1)}_{[2,2] (-2,2)} (\sigma_1, \sigma_2) 
+ 
\mathcal{P}^{(0)}_{[1,2] (0,2)} (\sigma_1, \sigma_2) r_{1 [1]}^{(1)} (\sigma_1) 
+ 
\mathcal{P}^{(0)}_{[2,1] (-2,0)} (\sigma_1, \sigma_2) r_{1 [1]}^{(1)} (\sigma_2)
\, .
\end{align}
In particular, $\mathcal{W}_{[2,2] (-2,2)} = \mathcal{W}_{\rm \bar{f}\bar{F}|Ff} + O (g^6)$, at leading order 
\begin{align}
\mathcal{W}^{(0)}_{[2,2] (-2,2)}
&
=
\mathcal{W}^{(0)}_{\rm \bar{f}\bar{F}|Ff}
=
- \pi^2 \int \frac{d u_2}{2 \pi} \frac{d v_1}{2 \pi} \frac{{\rm e}^{2 i u_2 \sigma_1 + 2 i v_1 \sigma_2}}{\sinh (\pi u_2) \sinh (\pi v_1)}
\frac{(u_2 - i) (v_1 - i) \Gamma (1 -  i u_2 - i v_1)}{\Gamma (1 - i u_2) \Gamma (1 -  i v_1)}
\nonumber\\
&
=
- \frac{ {\rm e}^{6 \sigma_1} 
+ 3 {\rm e}^{4 \sigma_1 + 2 \sigma_2} 
+ 5 {\rm e}^{6 \sigma_1 + 2 \sigma_2} 
+ 6 {\rm e}^{4 \sigma_1 + 4 \sigma_2} 
+ 10 {\rm e}^{6 \sigma_1 + 4 \sigma_2} 
+ 3 {\rm e}^{6 \sigma_1 + 6 \sigma_2} 
+ (\sigma_1 \leftrightarrow \sigma_2)
}{
(1 + {\rm e}^{2 \sigma_1})^2
(1 + {\rm e}^{2 \sigma_2})^2
({\rm e}^{2 \sigma_1} + {\rm e}^{2 \sigma_2} + {\rm e}^{2 \sigma_1 + 2 \sigma_2})^2
}
\, ,
\end{align}
and a very lengthy expression for the one-loop amplitude (see the notebook).

\subsubsection{Fermion-gluon states}

Next, we address the $\chi_1^3 \chi_4$ component and start by exploring its $\mathcal{W}_{[2,2] (3,3)}$ contribution. The latter is determined by the
gluon-fermion flux-tube excitations created at the bottom and absorbed by the top of the heptagon,
\begin{align}
\mathcal{W}_{[2,2](3,3)} 
= \mathcal{W}_{\Psi {\rm g}| \Psi {\rm g}}
=
\int 
&
d \mu_\Psi (u_1) d \mu_{\rm g} (u_2) \, i x [u_1] \, \sqrt{x^+ [u_2] x^- [u_2]}
F_{\Psi{\rm g}} (0 |u_1, u_2) 
\\
&
\times P_{{\rm g}\Psi | \Psi {\rm g}} (- u_2, - u_1 | v_2, v_1) \frac{F_{{\rm g}\Psi} ( - v_1, - v_2 | 0)}{\sqrt{x^+ [v_1] x^- [v_1]}}
d \mu_\Psi (v_2) d \mu_{\rm g} (v_1)
\, . \nonumber
\end{align}
Here the production and annihilation form factors for the flux-tube pair are given in Eqs.\ \re{PsiPsiFF} and \re{PsiGproductionFF}, respectively,
while the $P_{{\rm g}\Psi | \Psi {\rm g}}$ is determined by the factorized expression \re{2to2Pentagon}. As before, $\mathcal{W}_{\Psi {\rm g}| \Psi {\rm g}}$
can be decomposed in four terms
\begin{align}
\mathcal{W}_{\Psi {\rm g}| \Psi {\rm g}}
=
\mathcal{W}_{{\rm fg}|{\rm fg}}
+
\mathcal{W}_{{\rm Fg}|{\rm fg}}
+
\mathcal{W}_{{\rm fg}|{\rm Fg}}
+
\mathcal{W}_{{\rm Fg}|{\rm Fg}}
\, ,
\end{align}
where the first one reads
\begin{align}
\mathcal{W}_{{\rm fg}|{\rm fg}}
=
\frac{1}{g^2}
\int 
&
d \mu_{\rm fg} (u_2) 
\sqrt{x^+ [u_2] x^- [u_2]} P_{\rm g|f} ( - u_2| v_1^-) 
\\
&
\times
P_{\rm g|g} (- u_2 | v_1) P_{\rm f|f} (- u_2^- | v_1^-)  P_{\rm f|g} ( - u_2^- | v_1) 
\frac{x [v_1^-]}{\sqrt{x^+ [v_1] x^- [v_1]}} d \mu_{\rm fg} (v_1)
\, . \nonumber
\end{align}
It accommodates the tree amplitude that reads
\begin{align}
\mathcal{W}^{(0)}_{[2,2](3,3)} 
= 
\mathcal{W}^{(0)}_{{\rm fg}|{\rm fg}}
&
=
\pi^2 \int \frac{d u_2}{2 \pi} \frac{d v_1}{2 \pi} \frac{{\rm e}^{2 i u_2 \sigma_1 + 2 i v_1 \sigma_2}}{\cosh (\pi u_2) \cosh (\pi v_1)}
\frac{(u_2 + v_1 - i) \Gamma (-  i u_2 - i v_1)}{(v_1 + \ft{i}{2}) \Gamma (- \ft12 - i u_2) \Gamma ( -\ft12 -  i v_1)}
\nonumber\\
&
=
- \frac{ {\rm e}^{3 \sigma_1 + \sigma_2} 
( {\rm e}^{2 \sigma_1} + {\rm e}^{2 \sigma_2} - {\rm e}^{2 \sigma_1 + 2 \sigma_2} - 2 {\rm e}^{4 \sigma_1 + 2 \sigma_2} )
}{
(1 + {\rm e}^{2 \sigma_1})^2
({\rm e}^{2 \sigma_1} + {\rm e}^{2 \sigma_2} + {\rm e}^{2 \sigma_1 + 2 \sigma_2})^3
}
\, .
\end{align}
A simple counting argument reveals that both $\mathcal{W}_{{\rm fg}|{\rm Fg}}$ and $\mathcal{W}_{{\rm Fg}|{\rm Fg}}$ are of oder
$O (g^6)$ and thus start at two-loop order only, while $\mathcal{W}_{{\rm Fg}|{\rm fg}}$ contributes at one-loop already and has to be
accounted for. It yields
\begin{align}
\mathcal{W}^{(1)}_{{\rm Fg}|{\rm fg}}
=
- i \pi^4 \int 
\frac{d u_1}{2 \pi} \frac{d u_2}{2 \pi} \frac{d v_1}{2 \pi} 
&
\frac{{\rm e}^{2 i ( u_1 + u_2) \sigma_1 + 2 i v_1 \sigma_2}}{\sinh(\pi u_1) \cosh (\pi u_2) \cosh (\pi v_1)}
\\
&
\times
\frac{(u_2 + v_1 - i) \Gamma (-  i u_2 - i v_1) \Gamma (\ft12 - i u_1 -  i v_1)
}{
\Gamma (1 - i u_1) \Gamma (\ft32 - i u_2) \Gamma ( -\ft12 -  i v_1)  \Gamma ( \ft32 -  i v_1)}
\nonumber
\, ,
\end{align}
and when added to the one-loop expansion to the $\mathcal{W}_{{\rm fg}|{\rm fg}}$ reproduces the subtracted ratio function
\begin{align}
\mathcal{W}^{(0)}_{[2,2](3,3)} (\sigma_1, \sigma_2)
&
=
\mathcal{P}^{(0)}_{[2,2](3,3)} (\sigma_1, \sigma_2)
\, , \\
\mathcal{W}^{(1)}_{[2,2] (3,3)}  (\sigma_1, \sigma_2)
&= \mathcal{P}^{(1)}_{[2,2] (3,3)} (\sigma_1, \sigma_2) 
+ 
\mathcal{P}^{(0)}_{[1,2] (1,3)} (\sigma_1, \sigma_2) r_{1 [1]}^{(1)} (\sigma_1) 
\nonumber\\
&
+
\mathcal{P}^{(0)}_{[2,1] (3,1)} (\sigma_1, \sigma_2) r_{1 [1]}^{(1)} (\sigma_2) 
+ 
\mathcal{P}^{(0)}_{[1,1] (1,1)} (\sigma_1, \sigma_2) r_{2 [2]}^{(1)} (\sigma_1, \sigma_2)
\, .
\end{align}

\subsubsection{Antifermion-hole states}

Let us finally analyze the transition of the antifermion-hole into the fermion-gluon pair. To this end, we consider the $\mathcal{W}_{[2,2] (-1,3)}$
contribution to the $\chi_1^3 \chi_4$ Grassmann component. Its relation to the ratio function can be found from the general formula \re{HeptagonWsubtracted} 
and gives to leading and next-to-leading orders
\begin{align}
\mathcal{W}^{(0)}_{[2,2] (-1,3)} (\sigma_1, \sigma_2)
&
= \mathcal{P}^{(0)}_{[2,2] (-1,3)} (\sigma_1, \sigma_2)
\, , \\
\mathcal{W}^{(1)}_{[2,2] (-1,3)}  (\sigma_1, \sigma_2)
&= \mathcal{P}^{(1)}_{[2,2] (-1,3)} (\sigma_1, \sigma_2) 
+ 
\mathcal{P}^{(0)}_{[1,2] (1,3)} (\sigma_1, \sigma_2) r_{1 [1]}^{(1)} (\sigma_1) 
+
\mathcal{P}^{(0)}_{[2,1] (-1,1)} (\sigma_1, \sigma_2) r_{1 [1]}^{(1)} (\sigma_2) 
\, .
\end{align}
In terms of the flux-tube excitations, $\mathcal{W}_{[2,2] (-1,3)}$ can be written as a sum of two contributions
\begin{align}
\mathcal{W}_{[2,2] (-1,3)} = \mathcal{W}_{\bar\Psi{\rm h}| \Psi {\rm g}} + \mathcal{W}_{\Psi{\rm \bar{g}}| \Psi {\rm g}} 
\, .
\end{align}
As in several cases analyzed before, only the first term in the sum induces a nonvanishing effects at lowest orders of perturbation theory and takes the form
\begin{align}
\mathcal{W}_{\bar\Psi{\rm h}| \Psi {\rm g}} 
=
\int 
&
d \mu_\Psi (u_1) d \mu_{\rm h} (u_2) \, F^{\rm 4}_{\bar\Psi{\rm h}} (0|u_1, u_2)
\\
&\times
\frac{g^2 P_{{\rm h}\bar\Psi|\Psi{\rm g}} (- u_2, - u_1| v_2, v_1)}{x^+ [v_1] x^- [v_1]} \frac{i \sqrt{x[v_2]}}{g}  F_{{\rm g} \Psi} (- v_1, - v_2| 0)
d \mu_{\rm g} (v_1) d \mu_{\Psi} (v_2)
\, . \nonumber
\end{align}
In fact, out of four contributions spawned by this expression $\mathcal{W}_{\bar\Psi{\rm h}| \Psi {\rm g}} = \mathcal{W}_{{\rm \bar{f}h}| {\rm fg}} + 
\mathcal{W}_{{\rm \bar{F}h}| {\rm fg}} + \mathcal{W}_{{\rm \bar{f}h}| {\rm Fg}} + \mathcal{W}_{{\rm \bar{F}h}| {\rm Fg}} $, only the one with both 
fermions belonging to the small fermion sheet accounts for the entire tree and one-loop amplitudes,
\begin{align}
\mathcal{W}_{{\rm \bar{f}h}| {\rm fg}} 
=
\frac{1}{g}
\int d \mu_{\rm hf} (u_2)
&
P_{{\rm h|f}} (- u_2 | v_1 - \ft{i}{2})
\frac{P_{\rm h|g} (- u_2 | v_1)}{\sqrt{x^+ [v_1] x^- [v_1]}}
\\
\times&
P_{\rm \bar{f}| f} (- u_2 + \ft{3i}{2}| v_1 - \ft{i}{2})
\frac{P_{\rm \bar{f}|g} (- u_2 + \ft{3i}{2} | v_1)}{\sqrt{x^+ [v_1] x^- [v_1]}}
\sqrt{ x^- [v_1]}
d \mu_{\rm gf} (v_1)
\, . \nonumber
\end{align}
Explicitly, it yields the tree level result
\begin{align}
\mathcal{W}^{(0)}_{{\rm \bar{f}h}| {\rm fg}} 
&
=
- i \pi^2 \int \frac{d u_2}{2 \pi} \frac{d v_1}{2 \pi} \frac{{\rm e}^{2 i u_2 \sigma_1 + 2 i v_1 \sigma_2}}{\cosh (\pi u_2) \cosh (\pi v_1)}
\frac{(u_2 - \ft{3 i}{2}) (v_1 - \ft{i}{2})\Gamma (1 -  i u_2 - i v_1)}{(v_1 + \ft{i}{2}) \Gamma (\ft12 - i u_2) \Gamma (\ft12 -  i v_1)}
\nonumber\\
&
=
\frac{ {\rm e}^{5 \sigma_1 + \sigma_2} 
( 2 {\rm e}^{2 \sigma_1} + {\rm e}^{4 \sigma_1} + 4 {\rm e}^{2 \sigma_2}  + 5 {\rm e}^{2 \sigma_1 + 2 \sigma_2} + {\rm e}^{4 \sigma_1 + 2 \sigma_2} )
}{
(1 + {\rm e}^{2 \sigma_1})^2
({\rm e}^{2 \sigma_1} + {\rm e}^{2 \sigma_2} + {\rm e}^{2 \sigma_1 + 2 \sigma_2})^3
}
\, ,
\end{align}
with the rest summarized in the accompanying file.

\subsection{A glimpse into higher twists: three-particle states}

To conclude the discussion of the heptagon, let us take a glance at multiparticle states with twist higher than two. A complete treatment requires analysis of 
pentagon transitions involving gluonic bound states paired with other flux-tube excitations. Therefore, let us content ourselves with a contribution that is insensitive 
to these (at least to lowest orders in 't Hooft coupling). We will consider twist-three contribution, i.e., ${\rm e}^{- 3 \tau_1 - \tau_2}$ to the $\chi_1^3 \chi_4$ Grassmann component of the super-Wilson 
loop. It reads in terms of the corresponding amplitudes
\begin{align}
\mathcal{W}^{(0)}_{[3,1] (-3,1)} (\sigma_1, \sigma_2)
&= \mathcal{P}^{(0)}_{[3,1] (-3,1)} (\sigma_1, \sigma_2)
\, ,\\
\mathcal{W}^{(1)}_{[3,1] (-3,1)}  (\sigma_1, \sigma_2)
&= \mathcal{P}^{(1)}_{[3,1] (-3,1)} (\sigma_1, \sigma_2) 
+ 
\mathcal{P}^{(0)}_{[2,1] (-1,1)} (\sigma_1, \sigma_2) r_{1 [1]}^{(1)} (\sigma_1) 
+ 
\mathcal{P}^{(0)}_{[1,1] (1,1)} (\sigma_1, \sigma_2) r_{1 [2]}^{(1)} (\sigma_1)
\, ,
\end{align}
and arises from the production of three antifermion flux-tube excitations in the {\bf 4} of SU(4), i.e., $\varepsilon^{ABCD} \bar\psi_B  \bar\psi_C  
\bar\psi_D$ that undergo a transition into a single fermion at the top of the heptagon,
\begin{align}
\mathcal{W}_{[3,1] (-3,1)} (\sigma_1, \sigma_2)
&
=
\frac{1}{3!}
\int d \mu_{\Psi} (u_1) d \mu_{\Psi} (u_2) d \mu_{\Psi} (u_3) d \mu_{\Psi} (v_1)
x [u_1] x [u_2] x [u_3] x [v_1] 
\nonumber\\
&\times
F^{\bf 4}_{\bar\Psi \bar\Psi \bar\Psi} (0 | u_1, u_2, u_3) P_{\bar\Psi \bar\Psi \bar\Psi | \Psi} (- u_3, - u_2, - u_1 | v_1)
\, .
\end{align}
Here the form factor for creation of three antifermions as well as the three-to-one pentagon transition both admit factorized forms
\begin{align}
&
F^{\bf 4}_{\bar\Psi \bar\Psi \bar\Psi} (0 | u_1, u_2, u_3) 
=
\prod_{i < j}^3
\frac{i}{(u_i - u_j + i) P_{\Psi | \Psi} (u_i | u_j)}
\, , \\
&
P_{\bar\Psi \bar\Psi \bar\Psi | \Psi} (- u_3, - u_2, - u_1 | v_1)
=
\frac{ \prod_{i = 1}^3 P_{\bar\Psi | \Psi} (- u_i | v_1)}{\prod_{i < j}^3 P_{\Psi | \Psi} (- u_i | - u_j)}
\end{align}
The portion of the integration contour $C$ that belongs to the small fermion sheet induces the leading contribution to the component of the Wilson
loops in question. Namely, two out of three antifermions are on the small fermion sheet with the remaining one belongs to the large one. Evaluating
the resulting integrals via the Cauchy theorem by picking up the residues in the rational prefactors in the particle creation form factor, we find
\begin{align}
\mathcal{W}_{[3,1] (-3,1)} (\sigma_1, \sigma_2)
&
=
\frac{1}{2}
\int d \mu_{\rm F} (u_1) \mu_{\rm f} (u_1 - i) \mu_{\rm f} (u_1 - 2 i)
\int d \mu_{\rm F} (v_1) \frac{x [u_1] x[v_1]}{x[u_1 - i] x[u_1 -2 i]}
\nonumber\\
&
\times
\frac{
P_{\rm \bar{f} | F} (- u_1 + 2 i| v_1) P_{\rm \bar{f} | F} (- u_1 + i| v_1) P_{\rm \bar{F} | F} (- u_1| v_1)
}{
[ P_{\rm f | f} (u_1 - 2 i| u_1 - i)]_\pm^2 [P_{\rm f | F} (u_1 - 2 i| u_1)]_\pm^2 [P_{\rm f | F} (u_1 - i| u_1) ]_\pm^2
}
\, ,
\end{align}
where we introduced a shorthand notation $[ P_{\rm p_1 | p_2} (u_1| u_2)]_\pm^2 \equiv P_{\rm p_1 | p_2} (u_1 | u_2) P_{\rm p_1 | p_2} (- u_1 | - u_2)$.
Expanding the integrand to the lowest two orders of perturbation theory, we immediately find an exact agreement with the superloop component
$\mathcal{W}_{[3,1] (-3,1)}$. For reference, we quote the leading order result
\begin{align}
\mathcal{W}_{[3,1] (-3,1)}^{(0)}
&
 (\sigma_1, \sigma_2)
=
- \frac{\pi^2}{2} \int \frac{d u_1}{2 \pi} \frac{d v_1}{2 \pi} \,
\frac{ {\rm e}^{2 i u_1 \sigma_1 + 2 i v_1 \sigma_2} }{\sinh (\pi u_1) \sinh (\pi v_1)}
\frac{(u_1 - i) (u_1 - 2 i)\Gamma (1 - i u_1 - i v_1)}{\Gamma (1 - i u_1) \Gamma (1 - i v_1)}
\\
&
= 
- 
\frac{
({\rm e}^{2 \sigma_1} + {\rm e}^{2 \sigma_2} + {\rm e}^{2 \sigma_1 + 2 \sigma_2})^3
+ 
3 {\rm e}^{2 \sigma_1 + 4 \sigma_2}
(1 + {\rm e}^{2 \sigma_1})
({\rm e}^{2 \sigma_1} + {\rm e}^{2 \sigma_2} + {\rm e}^{2 \sigma_1 + 2 \sigma_2})
+
 {\rm e}^{6 \sigma_1 + 2 \sigma_2}
}{(1 + {\rm e}^{2 \sigma_1})^3 (1 + {\rm e}^{2 \sigma_2}) ({\rm e}^{2 \sigma_1} + {\rm e}^{2 \sigma_2} + {\rm e}^{2 \sigma_1 + 2 \sigma_2})^3}
\, . \nonumber
\end{align}
The one-loop expression is too lengthy to be displayed here and can be found in the companion Mathematica notebook. At two loops and higher, 
the amplitude receives additive terms from other multi-particle contributions with the same quantum numbers, like two-gluon bound state accompanied 
by a fermion and two-gluon--fermion states. Their analysis goes beyond the scope of the present work and is deferred to a future publication.

\section{Octagon observable}
\label{OctagonSect}

The operator product expansion analysis of the octagon follows the same footsteps. The octagonal super Wilson loop is related in the same fashion to 
the eight-particle super-ratio function as for the heptagon \re{HeptagonWsubtracted},
\begin{align}
\label{OctagonWsubtracted}
\mathcal{W}_{8;n} = g^{2 n} \mathcal{P}_{8;n} W_8
\, ,
\end{align}
with the only difference that the bosonic loop $W_8$ receives more terms at each loop order in its perturbative expansion,
\begin{align}
W_8 &
=
1
+ g^2 \Big[
r_1 (\tau_1, \sigma_1, \phi_1) 
+ 
r_1 (\tau_2, \sigma_2, \phi_2) 
+ 
r_1 (\tau_3, \sigma_3, \phi_3) 
\\
&
+ 
r_2 (\tau_1, \tau_2, \sigma_1, \sigma_2, \phi_1, \phi_2) 
+ 
r_2 (\tau_2, \tau_3, \sigma_2, \sigma_3, \phi_2, \phi_3) 
+
r_3 (\tau_1, \tau_2, \tau_3, \sigma_1, \sigma_2, \sigma_3, \phi_1, \phi_2, \phi_3) 
\Big]
+ O (g^4)
\, . \nonumber
\end{align}
While the functions $r_1$ and $r_2$ are the same as in Section \ref{HeptagonSection} and were determined earlier in Eqs.\ \re{Wr1} and \re{Wr2}, respectively, 
the new ingredient $r_3$ depends on all three triplets of conformal cross ratios (see Appendix \ref{AppTwistors}) and reads to leading order in the OPE
\begin{align}
r_3 (\tau_1, \tau_2, \tau_3, \sigma_1, \sigma_2, \sigma_3, \phi_1, \phi_2, \phi_3) 
=
{\rm e}^{- \tau_1 - \tau_2 - \tau_3}
({\rm e}^{i \phi_1 + i \phi_2 + i \phi_3} + {\rm e}^{- i \phi_1 - i \phi_2 - i \phi_3})
r_{3 [3]} (\sigma_1, \sigma_2, \sigma_3)
+
\dots
\, ,
\end{align}
where the $\sigma$-dependent coefficient is represented by a consecutive sequence of gluonic pentagon transitions
\begin{align}
r_{3 [3]} (\sigma_1, \sigma_2, \sigma_3)
=
\frac{1}{g^2}
\int d \mu_{\rm g} (u) P_{\rm g|g} (- u| v)  d \mu_{\rm g} (v) P_{\rm g|g} (- v| w)  d \mu_{\rm g} (w)
\, .
\end{align}
The helicity-violating contribution sets in starting from two loops \cite{Basso:2013aha} and is thus irrelevant for our present discussion. 

Due to a plethora of various  contributions to the octagon OPE, let us discuss a few illustrative examples. The NMHV octagon admits the following 
Grassmann decomposition
\begin{align}
\mathcal{W}_{8;1} 
= 
\chi_1^2 \chi_4^2 
\bigg\{
&
{\rm e}^{- \tau_1 - \tau_2 - \tau_3}
\mathcal{W}_{[1,1,1](0,0,0)} 
\nonumber\\
+\, 
&
{\rm e}^{- 2 \tau_1 - \tau_2 - \tau_3} 
{\rm e}^{- i \phi_1} 
\mathcal{W}_{[2,1,1](- 2,0,0)} 
+
{\rm e}^{- \tau_1 - 2 \tau_2  - \tau_3} 
{\rm e}^{- i \phi_2} 
\mathcal{W}_{[1,2,1](0,- 2,0)} 
+ \dots
\bigg\}
\nonumber\\
+
\chi_1^3 \chi_4 
\bigg\{
&
{\rm e}^{- \tau_1 - \tau_2 - \tau_3} {\rm e}^{i \phi_1/2 + i \phi_2/2  + i \phi_3/2} 
\mathcal{W}_{[1,1,1](1,1,1)} 
\nonumber\\
+ \, 
& 
{\rm e}^{- 2 \tau_1 - \tau_2  - \tau_3} 
\left[
{\rm e}^{3 i \phi_1/2 + i \phi_2/2 + i \phi_3/2} 
\mathcal{W}_{[2,1,1](3,1,1)} 
+
{\rm e}^{- i \phi_1/2 + i \phi_2/2 +  i \phi_3/2} 
\mathcal{W}_{[2,1,1](-1,1,1)} 
\right]
+ \dots
\bigg\} 
\nonumber\\
+ \dots \, . \quad&
\end{align}
Below, we will provide in turn the flux-tube interpretation for corresponding coefficients to all orders in coupling and test them against available perturbative
data \cite{Bourjaily:2013mma}.

\subsection{Single-particle states}

The leading twist contributions to the components in question are
\begin{align}
\mathcal{W}_{[1,1,1](0,0,0)} 
&=
-
\int d \mu_{\rm h} (u) P_{\rm h|h} (- u| v) d \mu_{\rm h} (v) P_{\rm h|h} (- v| w) d \mu_{\rm h} (w)
\, , \\
\mathcal{W}_{[1,1,1](1,1,1)} 
&=
- i
\int d \mu_{\Psi} (u) x[u] \, P_{\Psi | \Psi} (- u| v) d \mu_{\Psi} (v) P_{\Psi | \Psi} (- v| w) d \mu_{\Psi} (w)
\, .
\end{align}
The first equation here was already discussed in Ref.\ \cite{Basso:2013aha}. In the second one, only the branch of the fermion on the large Riemann sheet induces 
a nonvanishing effect in weak coupling expansion, i.e., $\Psi = {\rm F}$.  At leading order, these read 
\begin{align}
\mathcal{W}^{(0)}_{[1,1,1](0,0,0)} 
&
=
\frac{- {\rm e}^{\sigma_1 + \sigma_2 + \sigma_3}
}{
{\rm e}^{2 \sigma_2}
+
{\rm e}^{2  \sigma_1 + 2 \sigma_2}
+
{\rm e}^{2  \sigma_1 + 2 \sigma_3}
+
{\rm e}^{2  \sigma_2 + 2 \sigma_3}
+
{\rm e}^{2  \sigma_1 + 2 \sigma_2 + 2 \sigma_3}
}
\, , \\
\mathcal{W}^{(0)}_{[1,1,1](1,1,1)} 
&=
\frac{{\rm e}^{2 \sigma_1 + 2 \sigma_3}
}{
(1 + {\rm e}^{2 \sigma_1})
(
{\rm e}^{2 \sigma_2}
+
{\rm e}^{2  \sigma_1 + 2 \sigma_2}
+
{\rm e}^{2  \sigma_1 + 2 \sigma_3}
+
{\rm e}^{2  \sigma_2 + 2 \sigma_3}
+
{\rm e}^{2  \sigma_1 + 2 \sigma_2 + 2 \sigma_3}
)}
\, .
\end{align}
Together with subleading corrections in $g^2$, calculated with expressions provided in the Appendix \ref{OneLoopPentagonsApp}, they agree with the one-loop ratio 
function (see the attached file).

\subsection{Two-particle states}

Let us analyze now twist-two contributions.

\subsubsection{Two-(anti)fermion states}

Starting with $\mathcal{W}_{[2,1,1](- 2,0,0)}$, the latter is determined by the sum of two twist-two components created at the bottom of the loop,
$\mathcal{W}_{[2,1,1](- 2,0,0)} = \mathcal{W}_{\bar\Psi\bar\Psi | {\rm h} | {\rm h}} +  \mathcal{W}_{{\rm \bar{g}h} | {\rm h} | {\rm h}}$.
However, as in the case of the heptagon discussed at the end of section \ref{Tw2Tw1PsiBarPsiBarSection}, only the first one
induces the leading two orders of perturbation theory,
\begin{align}
\mathcal{W}_{\bar\Psi\bar\Psi | {\rm h} | {\rm h}}
=
\int 
d \mu_{\Psi} (u_1) d \mu_{\Psi} (u_2)
\frac{\sqrt{x[u_1] x[u_2]}}{g^2}
&
F^{\bf 6}_{\Psi\Psi} (0 | u_1, u_2)
P_{\bar\Psi\bar\Psi | {\rm h}} (- u_2, - u_1| v_1) 
\\
\times
& 
d \mu_{\rm h} (v_1) P_{\rm h|h} (- v_1| w_1) d \mu_{\rm h} (w_1)
\, , \nonumber 
\end{align}
in particular, in the kinematics when one of the antifermions in the pair belongs to the small sheet while the other one to the large one,
\begin{align}
\mathcal{W}_{{\rm \bar{f}\bar{F}} | {\rm h} | {\rm h}}
=
\int d \mu_{\rm fF} (u_2) \frac{\sqrt{x[u_2 - i]/x[u_2]}}{g^2}
P_{\rm f|h} (- u_2 + i | v_1) P_{\rm F|h} (- u_2 | v_1) d \mu_{\rm h} (v_1) P_{\rm h|h} (- v_1| w_1) d \mu_{\rm h} (w_1)
\, .
\end{align}
At leading order, we obtain
\begin{align}
\mathcal{W}^{(0)}_{{\rm fF} | {\rm h} | {\rm h}}
&
= \pi^3
\int \frac{du_2}{2 \pi}  \frac{dv_1}{2 \pi}  \frac{dw_1}{2 \pi} 
\frac{{\rm e}^{2 i u_2 \sigma_1 + 2 i v_1 \sigma_2 + 2 i w_1 \sigma_3}}{\sinh(\pi u_2) \cosh (\pi v_1) \cosh (\pi w_1)}
\frac{(u_2 - i) \Gamma (\ft12 - i u_2 - i v_1) \Gamma (- i v_1 - i w_1)}{\Gamma (1 - i u_2) \Gamma^2 (\ft12 - i v_1) \Gamma (\ft12 - i w_1)}
\nonumber\\
&=
-
\frac{{\rm e}^{5 \sigma_2 + \sigma_3} (1 + {\rm e}^{2 \sigma_3})^2
}{
({\rm e}^{2 \sigma_2} + {\rm e}^{2 \sigma_3} + {\rm e}^{2 \sigma_2 + 2 \sigma_3} )
(
{\rm e}^{2 \sigma_2}
+
{\rm e}^{2  \sigma_1 + 2 \sigma_2}
+
{\rm e}^{2  \sigma_1 + 2 \sigma_3}
+
{\rm e}^{2  \sigma_2 + 2 \sigma_3}
+
{\rm e}^{2  \sigma_1 + 2 \sigma_2 + 2 \sigma_3}
)^2}
\, .
\end{align}
It agrees with the corresponding ratio function $\mathcal{P}^{(0)}_{[2,1,1](- 2,0,0)}$ along with its first subleading term $\mathcal{P}^{(1)}_{[2,1,1](- 2,0,0)}$,
\begin{align}
\label{PtoW211-0}
\mathcal{W}^{(0)}_{[2,1,1] (- 2,0,0)} (\sigma_1, \sigma_2, \sigma_3)
&= \mathcal{P}^{(0)}_{[2,1,1] (- 2,0,0)} (\sigma_1, \sigma_2, \sigma_3)
\, ,\\
\label{PtoW211-1}
\mathcal{W}^{(1)}_{[2,1,1] (- 2,0,0)}  (\sigma_1, \sigma_2, \sigma_3)
&= \mathcal{P}^{(1)}_{[2,1,1] (- 2,0,0)} (\sigma_1, \sigma_2, \sigma_3) 
+ 
\mathcal{P}^{(0)}_{[1,1,1] (0,0,0)} (\sigma_1, \sigma_2,  \sigma_3) r_{1 [1]}^{(1)} (\sigma_1) 
\, ,
\end{align}
as shown in the notebook.

\subsubsection{Antifermion-hole states}

Turning to the $\chi_1^3 \chi_4$ component, we consider the antifermion-hole state first. It follows the analysis in Sect.\ \ref{AntiFermionHoleHeptagonSection}, 
so we will be brief here. As for the heptagon, the lowest two orders in perturbation theory for $\mathcal{W}_{[2,1,1](-1,1,1)}$ are governed by the antifermion-hole
pair in the initial state $\mathcal{W}_{\bar{\Psi}{\rm h}| \Psi | \Psi}$, i.e., $\mathcal{W}_{[2,1,1](-1,1,1)} = \mathcal{W}_{\bar{\Psi}{\rm h}| \Psi | \Psi} + O(g^6)$, where
\begin{align}
\mathcal{W}_{\bar{\Psi}{\rm h}| \Psi}
&
=
\int d \mu_\Psi (u_1)\, d \mu_{\rm h} (u_2) \, F^{\bf 4}_{\bar\Psi {\rm h}} (0| u_1, u_2)
P_{{\rm h} \bar\Psi | \Psi} (- u_2, - u_1 | v_1) \frac{i \sqrt{x[v_1]}}{g^2}  \, d \mu_\Psi (v_1)
P_{\Psi | \Psi} (- v_1 | w_1)
\, d \mu_\Psi (w_1)
\, ,
\end{align}
and more precisely by the contribution from the small fermion sheet in the initial state
\begin{align}
\mathcal{W}_{\bar{\rm f}{\rm h}| {\rm F} | {\rm F}}
=
\int
d \mu_{\rm fh} (u_2) P_{\rm \bar{f}|F} (- u_2 + \ft{3 i}{2}| v_1)P_{\rm h|F} (- u_2 | v_1) 
\frac{\sqrt{x[v_1]}}{g} 
d \mu_{\rm F} (v_1) P_{\rm F|F} (- v_1 | w_1) d \mu_{\rm F} (w_1)
\, ,
\end{align}
Employing its perturbative expansion summarized in Appendix \ref{OneLoopPentagonsApp}, we find
\begin{align}
&
\mathcal{W}^{(0)}_{[2,1,1](-1,1,1)}
\nonumber\\
&\quad
= i \pi^3
\int \frac{du_2}{2 \pi}  \frac{dv_1}{2 \pi}  \frac{dw_1}{2 \pi} 
\frac{{\rm e}^{2 i u_2 \sigma_1 + 2 i v_1 \sigma_2 + 2 i w_1 \sigma_3}}{\cosh(\pi u_2) \sinh (\pi v_1) \sinh (\pi w_1)}
\frac{(u_2 - \ft{3i}{2}) \Gamma (\ft12 - i u_2 - i v_1) \Gamma (- i v_1 - i w_1)}{\Gamma (\ft{1}{2} - i u_2) \Gamma (1 - i v_1) \Gamma (- i v_1) \Gamma (1 - i w_1)}
\nonumber\\
&\quad
=
\frac{
{\rm e}^{3 \sigma_1 + 2  \sigma_3}
(3 + {\rm e}^{2 \sigma_1})
(
{\rm e}^{2 \sigma_2}
+
{\rm e}^{2  \sigma_1 + 2 \sigma_2}
+
{\rm e}^{2  \sigma_1 + 2 \sigma_3}
+
{\rm e}^{2  \sigma_2 + 2 \sigma_3}
+
{\rm e}^{2  \sigma_1 + 2 \sigma_2 + 2 \sigma_3}
)
-
{\rm e}^{5 \sigma_1 + 4  \sigma_3}
}{
(1 + {\rm e}^{2 \sigma_1})^2
(
{\rm e}^{2 \sigma_2}
+
{\rm e}^{2  \sigma_1 + 2 \sigma_2}
+
{\rm e}^{2  \sigma_1 + 2 \sigma_3}
+
{\rm e}^{2  \sigma_2 + 2 \sigma_3}
+
{\rm e}^{2  \sigma_1 + 2 \sigma_2 + 2 \sigma_3}
)^2}
\, .
\end{align}
The agreement persists for $\mathcal{W}^{(1)}_{[2,1,1](-1,1,1)}$ at one loop order (see the ancillary file) with the subtracted ratio function defined
by Eq.\ \re{PtoW211-1}, where one obviously has to replace the helicity subscripts as follows $(-2,0,0) \to (-1,1,1)$.

\subsubsection{Fermion-gluon states}

We finish the analysis of the $\chi_1^3 \chi_4$ component by unravelling the structure of its $\mathcal{W}_{[2,1](3,1)}$ term. The tree and one-loop 
coefficients in the perturbative expansion of the latter are related to the ratio function via the relations \re{PtoW211-0}  and \re{PtoW211-1}, where the
weight $(2,0,0)$ gets substituted by $(3,1,1)$. The analysis of quantum numbers suggests that this component is induced by the emission of the 
gluon-fermion pair at the bottom of the hexagon and absorption of a single fermion at the top. Thus its all-order expression reads
\begin{align}
\mathcal{W}_{[2,1,1](3,1,1)} 
=
\mathcal{W}_{\Psi{\rm g}|\Psi|\Psi}
&
=
\int d \mu_\Psi (u_1) \, d \mu_{\rm g} (u_2) \, i x[u_1] \, \frac{\sqrt{x^+ [u_2] x^- [u_2]}}{g}
\\
&\times F_{\Psi {\rm g}} (0| u_1, u_2)
P_{{\rm g} \Psi | \Psi} (- u_2, - u_1 | v_1) \, d \mu_\Psi (v_1)
P_{\Psi | \Psi} (- v_1 | w_1) \, d \mu_\Psi (w_1)
\, . \nonumber
\end{align}
The leading effects arise from the small fermion in the incoming state
\begin{align}
\mathcal{W}_{{\rm fg}| {\rm F}|{\rm F}}
=
\frac{1}{g}
\int
d \mu_{\rm fg} (u_2)  \, \sqrt{x^+ [u_2] x^- [u_2]} P_{\rm f|F} (- u_2 + \ft{i}{2}| v_1)P_{\rm g|F} (- u_2 | v_1) d \mu_{\rm F} (v_1) P_{\rm F|F} (- v_1 | w_1) d \mu_{\rm F} (w_1)
\, , 
\end{align}
The ratio function $\mathcal{W}_{[2,1,1](3,1,1)}$ with the leading term yielding explicitly
\begin{align}
&
\mathcal{W}^{(0)}_{[2,1,1](3,1,1)}
\nonumber\\
&\quad
=
\pi^3 \int \frac{d u_2}{2 \pi} \frac{d v_1}{2 \pi}  \frac{d w_1}{2 \pi} \frac{{\rm e}^{2 i u_2 \sigma_1 + 2 i v_1 \sigma_2 + 2 i w_1 \sigma_3}}{\cosh (\pi u_2) \sinh (\pi v_1) \sinh (\pi w_1)}
\frac{\Gamma (\ft12 -  i u_2 - i v_1) \Gamma (- v_1 - i w_1)}{\Gamma (- \ft12 - i u_2) \Gamma (1 -  i v_1) \Gamma (-  i v_1) \Gamma (1 -  i w_1)}
\nonumber\\
&\quad
=\frac{
{\rm e}^{3 \sigma_1 + 2  \sigma_3}
(
2 {\rm e}^{2 \sigma_2}
+
2 {\rm e}^{2  \sigma_1 + 2 \sigma_2}
+
{\rm e}^{2  \sigma_1 + 2 \sigma_3}
+
2 {\rm e}^{2  \sigma_2 + 2 \sigma_3}
+
2 {\rm e}^{2  \sigma_1 + 2 \sigma_2 + 2 \sigma_3}
)
}{
(1 + {\rm e}^{2 \sigma_1})^2
(
{\rm e}^{2 \sigma_2}
+
{\rm e}^{2  \sigma_1 + 2 \sigma_2}
+
{\rm e}^{2  \sigma_1 + 2 \sigma_3}
+
{\rm e}^{2  \sigma_2 + 2 \sigma_3}
+
{\rm e}^{2  \sigma_1 + 2 \sigma_2 + 2 \sigma_3}
)^2}
\, . 
\end{align}
The next-to-leading order in 't Hooft coupling for the ratio function coincides with the OPE prediction as well.

\section{Conclusions}

In this paper, we constructed the OPE for higher polygons (heptagons and octagons) within the integrability-based pentagon approach.
We considered the Grassmann degree-four components of the null polygonal Wilson loops which are dual to NMHV ratio function in maximally supersymmetric
Yang-Mills theory. The goal of this consideration was twofold. First, we tested the factorization hypothesis for multiparticle transitions in terms of single-particle
ones which is rooted in the integrability of the flux-tube dynamics. Currently, the fact that the bootstrap equations obeyed by fermionic excitations are nonlinear
in nature obscures the derivation of the factorized form for these transitions. Second, we verified the correctness of the charged single-particle pentagons derived 
from a set of postulated axioms in previous studies. Explicit perturbative data on scattering amplitudes involving more than six particles is rather scarce. Currently,
the only available source for an arbitrary number of legs is the one-loop calculation of Ref.\  \cite{Bourjaily:2013mma} cast in the form of a Mathematica routine. 
Making heavy use of the latter, both items on our agenda received positive confirmation. We observed that in all cases considered, two-particle contributions
involving at least one fermion induced tree-level amplitudes when the latter belonged to the small-fermion sheet thus acting as a supersymmetry transformation
on the accompanying flux-tube excitation. Compared to the previously analyzed NMHV hexagon, for higher polygons the onset of genuine two-particle states 
was lowered from two- to one-loop order exhibiting their stronger sensitivity to higher twist components of the flux-tube wave functions.

Having tested all charged pentagons, one can immediately generate results at any order of perturbation theory (or at finite coupling). A natural next step is to 
analyze the behavior of various components at strong coupling. Also there is no difficulty of principle to consider even higher polygons as well as any helicity 
configurations of incoming particles at higher twists. Currently, the only missing blocks on the way of achieving this for all possible components, are the charged 
pentagons involving the gauge field bound states undergoing transitions into other flux-tube excitations. This question is currently under study and results will be 
discussed elsewhere.

\section*{Acknowledgments}

We are grateful to Jacob Bourjaily, Lance Dixon and Jaroslav Trnka for instructive discussions and correspondence. We would like to thank Benjamin Basso
for informing us about a forthcoming work \cite{BasCaeCorSevVie15} where similar questions are addressed. This work was supported by the U.S. 
National Science Foundation under the grants PHY-1068286 and PHY-1403891.

\appendix

\setcounter{section}{0}
\setcounter{equation}{0}
\renewcommand{\theequation}{\Alph{section}.\arabic{equation}}

\section{Reference polygons}
\label{AppTwistors}

\begin{figure}[t]
\begin{center}
\mbox{
\begin{picture}(0,190)(215,0)
\put(0,-130){\insertfig{15}{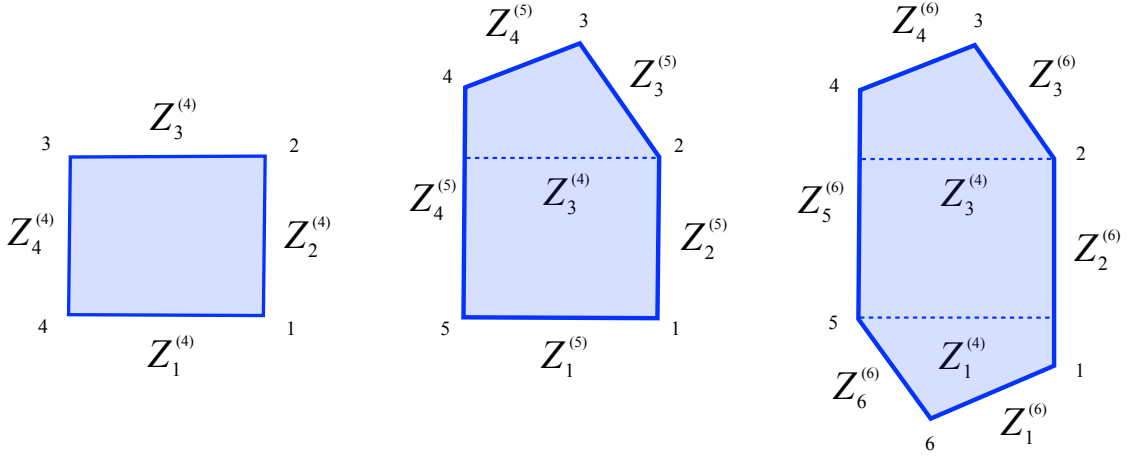}}
\end{picture}
}
\end{center}
\caption{ \label{PentagonFromSquare} Recursive construction of reference pentagons and hexagons from a square.}
\end{figure}

In this appendix we will outline a recursive construction of reference polygons which are used in the main body of the paper. We start from a 
reference square shown in (leftmost panel in) Fig.~\ref{PentagonFromSquare} that is defined by \cite{Basso:2013aha} 
\begin{align}
Z^{(4)}_1 = (0,0,1,0)
\, , \qquad
Z^{(4)}_2 = (1,0,0,0)
\, , \qquad
Z^{(4)}_3 = (0,0,0,1)
\, , \qquad
Z^{(4)}_4 = (0,1,0,0)
\, .
\end{align}
The latter is invariant under the conformal transformation
\begin{align}
Z^{(4)}_j = Z^{(4)}_j \cdot M (\tau, \sigma, \phi)
\end{align}
where the equality sign stands for ``equal up to rescaling'', with
\begin{align}
\label{MforRefSquare1}
M  (\tau, \sigma, \phi) = \left(
\begin{array}{cccc}
{\rm e}^{\sigma - i \phi/2} & & & \\
& {\rm e}^{-\sigma - i \phi/2} & & \\
& & {\rm e}^{\tau + i \phi/2} & \\
& & & {\rm e}^{- \tau + i \phi/2}
\end{array}
\right)
\, .
\end{align}
This matrix parametrizes the three conformal symmetries of the square that play a crucial role in the OPE framework \cite{Alday:2010ku}.

Adding two more twistor lines on top of the square as shown by the middle graph in Fig.~\ref{PentagonFromSquare}, one can construct a reference pentagon. As 
one can see, three out of five sides of the latter coincide with the square
\begin{align}
Z^{(5)}_1 = Z^{(4)}_1
\, , \qquad
Z^{(5)}_2 = Z^{(4)}_2
\, , \qquad
 Z^{(5)}_5 = Z^{(4)}_4
\, .
\end{align}
While the components of the remaining two, $Z^{(5)}_3$ and $Z^{(5)}_4$, are determined by the intersection and space-like interval conditions. Namely, the condition of
intersection of three twistor lines in the same point $X_2$,
\begin{align}
X_2 
\equiv
Z^{(4)}_2 \wedge Z^{(4)}_3 = Z^{(4)}_2 \wedge Z^{(5)}_3 = Z^{(4)}_3 \wedge Z^{(5)}_3
\end{align}
uniquely fixes $Z^{(5)}_3$ to be
\begin{align}
Z^{(5)}_3 = (-1,0,0,1)
\, .
\end{align}
The space-like nature of the intervals $(X^{(5)}_{13})^2$, $(X^{(5)}_{24})^2$ and $(X^{(5)}_{35})^2$, allows one to find (up to an overall scale) the last twistor
\begin{align}
Z^{(5)}_4 = (0,1,-1,1)
\, .
\end{align}

\begin{figure}[t]
\begin{center}
\mbox{
\begin{picture}(0,220)(160,0)
\put(0,-100){\insertfig{15}{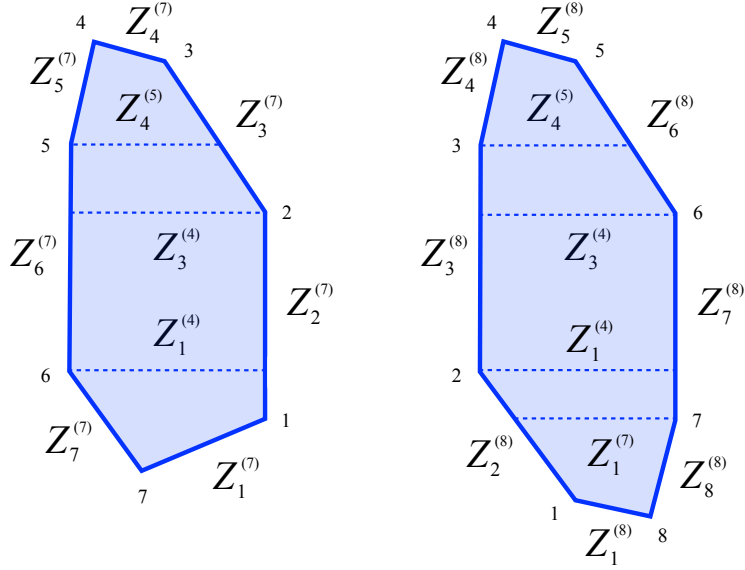}}
\end{picture}
}
\end{center}
\caption{ \label{OctagonFromHeptagon} Tessellation of heptagons and octagons.}
\end{figure}

Higher polygons are constructed accordingly by successively extending either the bottom or the top portions of the preceding polygons. The hexagon is demonstrated 
in the leftmost panel of Fig.~\ref{PentagonFromSquare} is encoded by the following twistors
\begin{align}
Z^{(6)}_1 &= (1,0,1,1)
\, , 
&
Z^{(6)}_2 
&
= (1,0,0,0)
\, , 
&
Z^{(6)}_3 
&
= (-1,0,0,1)
\, , \\
Z^{(6)}_4 
&= (0,1,-1,1)
\, ,
&
Z^{(6)}_5 
&
= (0,1,0,0)
\, , 
&
Z^{(6)}_6 
&
= (0,1,1,0)
\, . 
\end{align}
The heptagon and octagon are displayed in Fig.\ \ref{OctagonFromHeptagon}. Noticed that we flipped the assignment of twistors for the octagon
to have all polygons parametrized in the same fashion. The corresponding reference twistors are
\begin{align}
Z^{(7)}_1 &= (1,0,1,1)
\, , 
&
Z^{(7)}_2 
&
= (1,0,0,0)
\, , 
&
Z^{(7)}_3 
&
= (-1,0,0,1)
\, , \qquad
Z^{(7)}_4 
= (-1,1,-1,3)
\, , \nonumber\\
Z^{(7)}_5 
&= (0,2,-1,1)
\, ,
&
Z^{(7)}_6
&
= (0,1,0,0)
\, , 
&
Z^{(7)}_7
&
= (0,1,1,0)
\, ,
\end{align}
and 
\begin{align}
Z^{(8)}_1 &= (1,1,3,1)
\, , 
&
Z^{(8)}_2 
&
= (0,1,1,0)
\, , 
&
Z^{(8)}_3 
&
= (0,1,0,0)
\, , 
&
Z^{(8)}_4 
&
= (0,2,-1,1)
\, , \nonumber\\
Z^{(8)}_5 
&= (-1,1,-1,3)
\, ,
&
Z^{(8)}_6
&
= (-1,0,0,1)
\, , 
&
Z^{(8)}_7
&
= (1,0,0,0)
\, ,
&
Z^{(8)}_8 
&
= (2,0,1,1)
\, , 
\end{align}
respectively.

Notice that this construction provides a natural tessellation of null polygons: they are divided in a series of pentagon transitions that overlap on
intermediate null squares. To encode all inequivalent polygons we will apply conformal symmetries of these middle squares on all twistors above
or below them. All hexagons are then defined by the set
\begin{align}
\bit{Z}^{(6)} = \{ Z^{(6)}_1 \cdot M (\tau, \sigma, \phi) \, , \  Z^{(6)}_2 \, , \ 
Z^{(6)}_3 \, ,  \ Z^{(6)}_4 \, , \
Z^{(6)}_5 \, , \  Z^{(6)}_6  \cdot M (\tau, \sigma, \phi) 
\}
\, .
\end{align}
To define heptagons, while the bottom middle square is invariant under the same transformation $M$ as defined in Eq.\ \re{MforRefSquare1}, the top 
middle square is conformally invariant with respect to the matrix multiplication with
\begin{align}
\label{MforRefSquare2}
M^\prime  (\tau, \sigma, \phi) 
= 
\left(
\begin{array}{cccc}
{\rm e}^{- \sigma - i \phi/2} & & & - {\rm e}^{- \sigma - i \phi/2} + {\rm e}^{\tau + i \phi/2}  \\
& {\rm e}^{\sigma - i \phi/2} & & \\
& {\rm e}^{\sigma - i \phi/2} -  {\rm e}^{- \tau + i \phi/2} &  {\rm e}^{- \tau + i \phi/2} &  {\rm e}^{\tau + i \phi/2} -  {\rm e}^{- \tau + i \phi/2} \\
& & & {\rm e}^{\tau + i \phi/2}
\end{array}
\right)
\, .
\end{align}
Then all heptagons are parametrized by the set of twistors
\begin{align}
\label{HeptagonTwistors}
\bit{Z}^{(7)}
=
\{
&
Z^{(7)}_1 \cdot M (\tau_1, \sigma_1, \phi_1) \, , \
Z^{(7)}_2 \, , \
Z^{(7)}_3 \, , \
Z^{(7)}_4 
\cdot [ M^\prime (\tau_2, \sigma_2, \phi_2)]^{-1}
\, , \nonumber\\ 
&
Z^{(7)}_5 
\cdot [ M^\prime (\tau_2, \sigma_2, \phi_2)]^{-1}
\, , \
Z^{(7)}_6 \, , \ 
Z^{(7)}_7  \cdot M (\tau_1, \sigma_1, \phi_1) 
\}
\, .
\end{align}

To define the octagons, we have to find the symmetries of the bottom middle square. The latter is invariant with respect to the transformation matrix
\begin{align}
\label{MforRefSquare3}
M^{\prime\prime}  (\tau, \sigma, \phi) 
= 
\left(
\begin{array}{cccc}
{\rm e}^{- \sigma - i \phi/2} & & &  \\
& {\rm e}^{\sigma - i \phi/2} & {\rm e}^{\sigma - i \phi/2} - {\rm e}^{- \tau + i \phi/2} & \\
& &  {\rm e}^{- \tau + i \phi/2} &  \\
- {\rm e}^{- \sigma - i \phi/2} + {\rm e}^{\tau + i \phi/2}& & {\rm e}^{\tau + i \phi/2} -  {\rm e}^{- \tau + i \phi/2} & {\rm e}^{\tau + i \phi/2}
\end{array}
\right)
\, ,
\end{align}
such that the momentum twistors parametrizing all inequivalent octagons read
\begin{align}
\bit{Z}^{(8)}
=
\{
&
Z^{(8)}_1 \cdot M^{\prime\prime} (\tau_1, \sigma_1, \phi_1)  \cdot M (\tau_2, \sigma_2, \phi_2) 
\, , \ 
Z^{(8)}_2 \cdot M (\tau_2, \sigma_2, \phi_2) 
\, , \
Z^{(8)}_3 \, , \
Z^{(8)}_4 \cdot [ M^\prime (\tau_3, \sigma_3, \phi_3)]^{-1}
\, , \nonumber\\ 
&
Z^{(8)}_5 \cdot [ M^\prime (\tau_3, \sigma_3, \phi_3)]^{-1}
\, , \
Z^{(8)}_6
\, , \ 
Z^{(8)}_7 
\, , \
Z^{(8)}_8  \cdot M^{\prime\prime} (\tau_1, \sigma_1, \phi_1) \cdot M (\tau_2, \sigma_2, \phi_2) 
\}
\, .
\end{align}

\section{Pentagons, measures, energies and momenta}
\label{OneLoopPentagonsApp}

In this appendix, we summarize pentagon transitions for all single flux-tube excitations to one loop order. These obey the property
\begin{align}
P_{\rm p_1|p_2} (u_1|u_2) = P_{\rm \bar{p}_2|\bar{p}_1} (-u_2|-u_1) 
\, .
\end{align}
To simplify notations, we use the harmonic numbers of degree $r$, $H^{(r)}_{u}$ instead of Euler polygamma functions. We list below
the boson-boson, boson-fermion and fermion-fermion transitions, respectively.

\noindent Boson-boson pentagons \cite{Basso:2013aha,Belitsky:2014sla}:
\begin{align}
P_{\rm h|h} (u|v)
&
=
\frac{\Gamma (i u - i v)}{g^2 \, \Gamma (\ft12 + i u) \Gamma (\ft12 - i v)}
\\
&
+
\frac{\Gamma (i u - i v)}{\Gamma (\ft12 + i u) \Gamma (\ft12 - i v)}
\bigg[
H_{-1/2+iu} H_{-1/2+iv} + H_{-1/2 - i u} H_{-1/2 + iv}
\nonumber\\
&
- H_{-1/2 + iu} H_{-1/2 - iv} + H_{-1/2 - i u} H_{-1/2 - i v} - H^{(2)}_{-1/2+i u} - H^{(2)}_{-1/2 - i v}
\bigg]
+ 
O (g^2)
\, , \nonumber\\
P_{\rm h|g} (u|v)
&
=
\frac{\Gamma (1 + i u - i v)}{g \, \Gamma (\ft12 + i u) \Gamma (\ft12 - i v)} \sqrt{ \frac{v^-}{v^+} }
\\
&
+
\frac{g \Gamma (1 + i u - i v)}{\Gamma (\ft12 + i u) \Gamma (\ft12 - i v)} \sqrt{ \frac{v^-}{v^+} }
\bigg[
H_{-1/2 + i u} H_{-1/2 + i v} + H_{-1/2 - i u} H_{1/2 + i v} 
\nonumber\\
&
- H_{-1/2 + i u} H_{-1/2 - i v} + H_{-1/2 - i u} H_{1/2 - i v} -   H^{(2)}_{-1/2 + i u} - H^{(2)}_{-1/2 - i v} - \frac{i v}{(v^+ v^-)^2}
\bigg] 
+ O (g^3)
\, , \nonumber\\
P_{\rm g|g} (u|v)
&
=
- \frac{\Gamma (i u - i v)}{g^2 \Gamma (- \ft12 + i u) \Gamma (- \ft12 - i v)}
\\
&
-
\frac{\Gamma (i u - i v)}{2 \Gamma (- \ft12 + i u) \Gamma (- \ft12 - i v)}
\bigg[
8 \zeta_2 + \frac{1 - 4 u^2}{2 (u^+ u^-)^2} + \frac{1}{(v^+ v^-)^2} - \frac{1+ 4 u^2 + 8 u v}{2 u^+ u^- v^+ v^-}
\nonumber\\
&
+
\left(H_{1/2- i u} + H_{1/2 + i u} - i \pi \tanh (\pi u)\right) \left( H_{1/2 - i v} + H_{1/2 + i v} + i \pi \tanh (\pi v) \right)
\nonumber\\
&
+
H^{(2)}_{1/2 - iu} - H^{(2)}_{1/2 + i u} 
- 
H^{(2)}_{1/2 - iv} + H^{(2)}_{1/2 + i v}
- \pi^2 \tanh^2 (\pi u) - \pi^2 \tanh^2 (\pi v) 
\nonumber\\
&\qquad\qquad\qquad\qquad\qquad\qquad
- 2 \pi^2 \tanh (\pi u) \tanh (\pi v)
\bigg]
+ 
O (g^2)
\, , \nonumber\\
P_{\rm \bar{g}|g} (u|v)
&
=
\frac{\Gamma (2 + i u - i v)}{\Gamma (\ft32 + i u) \Gamma (\ft32 - i v)}
\\
&
+
\frac{g^2 \Gamma (2 + i u - i v )}{2 \Gamma (\ft32 + i u) \Gamma (\ft32 - i v)}
\bigg[
8 \zeta_2 - \frac{1 - 4 u^2}{2 (u^+ u^-)^2} - \frac{1}{(v^+ v^-)^2} + \frac{1+ 4 u^2 + 8 u v}{2 u^+ u^- v^+ v^-}
\nonumber\\
&
+
\left(H_{1/2- i u} + H_{1/2 + i u} - i \pi \tanh (\pi u)\right) \left( H_{1/2 - i v} + H_{1/2 + i v} + i \pi \tanh (\pi v) \right)
\nonumber\\
&
+
H^{(2)}_{1/2 - iu} - H^{(2)}_{1/2 + i u} 
- 
H^{(2)}_{1/2 - iv} + H^{(2)}_{1/2 + i v}
- \pi^2 \tanh^2 (\pi u) - \pi^2 \tanh^2 (\pi v) 
\nonumber\\
&\qquad\qquad\qquad\qquad\qquad\qquad
- 2 \pi^2 \tanh (\pi u) \tanh (\pi v)
\bigg]
+ 
O (g^4)
\, . \nonumber
\end{align}

\noindent Boson-fermion pentagons \cite{Belitsky:2014lta}:

\begin{align}
P_{\rm h|F} (u|v)
&
=
\frac{\sqrt{v} \Gamma (\ft12 + i u - i v)}{g \Gamma (\ft12 + i u) \Gamma (1 - i v)}
\\
&
+
\frac{g \,\sqrt{v} \Gamma (\ft12 + i u - i v)}{\Gamma (\ft12 + i u) \Gamma (1 - i v)}
\bigg[
H_{-1/2 - i u} H_{- i v}
- H_{-1/2 + i u} H_{- 1 - i v} 
\nonumber\\
&\qquad\qquad\qquad\qquad\ 
+
H_{-1/2 - i u} H_{i v}
+
H_{-1/2 + i u} H_{i v}
-
H^{(2)}_{-1/2 + i u}
-
H^{(2)}_{- 1 -i v}
+
\frac{1}{2 v^2}
\bigg]
+ O (g^4)
\, , \nonumber\\
P_{\rm h|f} (u|v)
&
=
\frac{g}{\sqrt{v}} + \frac{g^3}{v^{5/2}} \left[ \frac{1}{2} - i v H_{-1/2 + i u} \right] + O (g^4)
\, , \\
P_{\rm g|F} (u|v)
&
=
\frac{v \Gamma (\ft12 + i u - i v)}{g \, \Gamma (\ft12 + i u) \Gamma (1 - i v)}  \sqrt{\frac{u^+}{u^-}}
\\
&
+
\frac{g \, v \Gamma (\ft12 + i u - i v)}{\Gamma (\ft12 + i u) \Gamma (1 - i v)}  \sqrt{\frac{u^+}{u^-}}
\bigg[
\frac{i u}{(u^+ u^-)^2}
+
H_{1/2 + i u} H_{i v}
+
H_{1/2 - i u} H_{- 1 + i v}
\nonumber\\
&\qquad\qquad\qquad\qquad\ 
-
H_{-1/2 + i u} H_{- 1 - i v}
+
H_{-1/2 - i u} H_{- 1 - i v}
- 
H^{(2)}_{-1/2 + i u} -  H^{(2)}_{ - 1 - i v}
\bigg]
+
O (g^3)
\, , \nonumber\\
P_{\rm g|f} (u|v)
&
= \frac{i g  (u - v + \ft{i}{2})}{v \sqrt{x^+ [u] x^- [u]}}
\bigg\{
1
+
\frac{g^2}{v} 
\bigg[
\frac{1}{v} - i H_{- 3/2 + i u}
\bigg]
\bigg\}
+
O (g^5)
\, , \nonumber\\
P_{\rm \bar{g}|F} (u|v)
&
=
\frac{\Gamma (\ft32 + i u - i v)}{g \, \Gamma (\ft12 + i u) \Gamma (1 - i v)} \sqrt{\frac{u^+}{u^-}}
\\
&
+ \frac{g \Gamma (\ft32 + i u - i v)}{\Gamma (\ft12 + i u) \Gamma (1 - i v)} \sqrt{\frac{u^+}{u^-}}
\bigg[
\frac{i u}{(u^+ u^-)^2}
+ H_{1/2 + i u} H_{- 1 + i v}
- H_{- 1/2 + i u} H_{- i v}
\nonumber\\
&\qquad\qquad\qquad\qquad\ 
+ H_{1/2 - i u} H_{i v}
+ H_{- 1/2 - i u} H_{- i v}
- H^{(2)}_{- 1/2 + i u} - H^{(2)}_{- i v}
\bigg] + O (g^3)
\, , \nonumber\\
P_{\rm \bar{g}|f} (u|v)
&
= \frac{i}{g} \sqrt{x^+[u] x^-[u]}
\bigg\{
1
-
\frac{i g^2}{v}  H_{1/2 + i u}
+
O (g^4)
\bigg\}
\, . 
\end{align}

\noindent Fermion-fermion pentagons \cite{Basso:2014koa}:

\begin{align}
P_{\rm F|F} (u|v) 
&=
\frac{\Gamma (i u - i v)}{g^2 \Gamma (i u) \Gamma (- i v)}
\\
&
+
\frac{\Gamma (i u - i v)}{\Gamma ( i u) \Gamma (- i v)}
\bigg[
- \frac{1}{u v} 
+ H_{i u} H_{i v}
- H_{i u} H_{ - 1 - i v}
\nonumber\\
&\qquad\qquad\qquad\quad\ 
+ H_{- i u} H_{i v}
+ H_{ - 1 - i u} H_{- i v}
- H_{-1 + i u}^{(2)} - H_{- 1 - i v}^{(2)} 
\bigg]
+
O (g^2)
\, , \nonumber\\
P_{\rm f|F} (u|v) 
&
= \frac{i}{u} + \frac{g^2}{u^2} \left[ \frac{i}{u} - H_{- 1 - i v} \right] + O (g^4)
\, , \\
P_{\rm f|f} (u|v) 
&
= \frac{i}{u -v} - \frac{i g^2}{u v (u-v)} + O (g^4)
\, , \\
P_{\rm \bar{F}|F} (u|v) 
&=
\frac{\Gamma (1 + i u - i v)}{\Gamma (1 + i u) \Gamma (1- i v)}
\\
&
+
\frac{g^2 \Gamma (1 + i u - i v)}{\Gamma (1 + i u) \Gamma (1- i v)}
\bigg[
H_{i u} H_{-1 + i v}
- H_{- 1 + i u} H_{- i v}
\nonumber\\
&\qquad\qquad\qquad\quad\ 
+ H_{- i u} H_{i v}
+ H_{- i u} H_{- i v}
- H_{i u}^{(2)} - H_{- i v}^{(2)} 
\bigg]
+ O (g^4)
\, , \nonumber\\
P_{\rm \bar{f}|F} (u|v) 
&
= 1 + \frac{i g^2}{u} H_{- i v} + O (g^4)
\, , \\
P_{\rm \bar{f}|f} (u|v) 
&
= 1 + O (g^4)
\, .
\end{align}

\noindent The one-loop single-particle and gluon bound state measures are \cite{Basso:2013aha,Basso:2014koa}
\begin{align}
\mu_{\rm g} (u)
&
= - \frac{g^2 \pi}{u^+ u^- \cosh (\pi u)}
\\
&
- \frac{g^4 \pi}{ 2 u^+ u^- \cosh (\pi u)}
\bigg[
- 
\left( H_{1/2 - i u} + H_{1/2 + i u} \right)^2
+
10 \zeta_2 - \frac{1 - 8 u^2}{(u^+ u^-)^2} 
-
\frac{18 \zeta_2}{\cosh^2 (\pi u)}
\bigg]
+
O (g^6)
\, , \nonumber\\
\mu_{\rm h} (u)
&
=
\frac{g^2 \pi}{\cosh (\pi u)}
\\
&
+
\frac{g^4 \pi}{\cosh (\pi u)}
\bigg[
-
H_{-1/2 - i u}^2 - H_{- 1/2 + i u}^2
+
2 \zeta_2 \big( 1 - 3 {\rm sech}^2 (\pi u) \big)
\bigg]
+
O (g^6)
\, , \nonumber\\
\mu_{\rm F} (u)
&
= 
\frac{g^2 \pi}{u \sinh(\pi u)}
\\
&
+
\frac{g^4 \pi}{u \sinh(\pi u)}
\bigg[
- H_{- i u}^2 - H_{i u}^2
+
\frac{1}{u^2} + \frac{\pi \coth (\pi u)}{u}
+
2 \zeta_2
+
\frac{\pi^2}{\sinh^2 (\pi u)}
\bigg]
+ O (g^6)
\, , \nonumber\\
\mu_{\rm f} (u)
&
= - 1 - \frac{g^2}{u^2} + O (g^4)
\, , \\
\mu_{D \rm g} (u)  
&
= 
\frac{g^2 \pi u}{(u^2 + 1) \sinh (\pi u)}
\\
&
+
\frac{g^4 \pi u}{(u^2 + 1) \sinh (\pi u)}
\bigg[
- H_{- i u}^2 - H_{i u}^2
- 2 
\frac{H_{- i u} + H_{i u}}{u^2 + 1}
-
\frac{1 + 6 u^2 - 3 u^4}{u^2 (u^2 + 1)^2}
+
2 \zeta_2
+
\frac{\pi^2}{\sinh^2 (\pi u)}
\bigg]
\nonumber\\
&
+ O (g^6)
\, , \nonumber
\end{align}
while the composite two-particle measures of a small fermion accompanying other flux-tube excitations read \cite{Belitsky:2014sla,Belitsky:2014lta}
\begin{align}
\mu_{\rm fF} (u) 
&
= \frac{\pi (u - i)}{\sinh(\pi u)}
\bigg\{
g^2
\\
&
+
g^4 
\left[
- H_{i u}^2 - H_{- i u}^2
+
2 \zeta_2  - \frac{1 + i \pi \coth (\pi u)}{u (u - i)} + \frac{\pi^2}{\sinh^2 (\pi u)} 
\right]
+ O(g^6)
\bigg\}
\, , \nonumber\\
\mu_{\rm hf} (u) 
&
=
\frac{\pi \left(u - \ft{3i}{2}\right)}{i \cosh(\pi u)}
\bigg\{ g^2
\\
&
+ g^4
\bigg[
- H_{-1/2-i u}^2 - H_{-1/2+i u}^2 + 2 \zeta_2 (1 - 3 \, {\rm sech}^2 (\pi u)) - \frac{\pi \tanh (\pi u)}{u - \ft{3 i}{2}}
\bigg]
+
O (g^6)
\bigg\}
\, , \nonumber\\
\mu_{\rm gf} (u) 
&
= 
\frac{\pi u^-}{i \cosh (\pi u)}
\bigg\{
g^2
\\
&+ g^4
\left[ 
- \ft{1}{2} \left( H_{1/2 - i u} + H_{1/2 + i u} \right)^2 - \frac{\pi  \left( 3 \pi u^- + \sinh (2 \pi u) \right)}{2 u^- \cosh^2 (\pi u)}
-
\frac{i u}{(u^+ u^-)^2} + 5 \zeta_2
\right]
+
O (g^6)
\bigg\}
\, . \nonumber
\end{align}

Finally, we quote the flux-tube dispersion relations to the required order \cite{Belitsky:2006en,Belitsky:2008mg,Basso:2010in}
\begin{align}
E_{\rm h}&=
1 + 2 g^2 \left( H_{- 1/2 - i u} + H_{-1/2 + i u} \right) + O (g^4)
\, , \quad
&
p_{\rm h} 
&
= 2 u - 2 g^2 \pi \tanh (\pi u) + O (g^4)
\, , \\
E_{\rm g}&=
1 + 2 g^2 \left( H_{1/2 - i u} + H_{1/2 + i u} \right) + O (g^4)
\, , \quad
&
p_{\rm g} 
&
= 2 u - 2 g^2 \pi \tanh (\pi u) + O (g^4)
\, , \\
E_{\rm F}&=
1 + 2 g^2 \left( H_{- i u} + H_{i u} \right) + O (g^4)
\, , \quad
&
p_{\rm F} 
&
= 2 u - 2 g^2 \pi \coth (\pi u) + O (g^4)
\, , \\
E_{\rm f}&= 1 + O (g^6)
\, , \quad
&
p_{\rm f} 
&
= 2 g^2 /u + O (g^4)
\, .
\end{align}
In the above equations as well as the main body of the paper, we used the definition of the Zhukowski variable
\begin{align}
\label{ZhukowskiVariable}
x[u] = \ft12 (u + \sqrt{u^2 - (2 g)^2})
\, ,
\end{align}
as well as the following conventions for shifted rapidities
\begin{align}
\label{ShiftedZhukowski}
u^\pm \equiv u \pm \frac{i}{2}
\, , \qquad
x^\pm [u] \equiv x[u^\pm]
\, .
\end{align}


\end{document}